\documentclass[article,10pt,twocolumn]{IEEEtran}
\usepackage{blindtext}
\usepackage[table]{xcolor}
\usepackage{tabularx}
\usepackage{comment}
\usepackage{multicol}
\usepackage{booktabs}
\usepackage{balance}
\usepackage{color, colortbl}

\usepackage[numbers, square, comma, sort&compress]{natbib}
\usepackage{nomencl}
\makeglossary
\usepackage{pdflscape}
\usepackage{soul}

\usepackage{enumitem}
\usepackage{cancel}
\usepackage[english]{babel}
\usepackage{multicol}
\usepackage{multirow}
\usepackage{amsmath}
\usepackage{amsfonts}
\usepackage{amssymb}
\usepackage{float}
\usepackage{stfloats}
\usepackage{subfigure}
\usepackage[hyphens]{url}
\usepackage[hidelinks]{hyperref}
\hypersetup{breaklinks=true}
\usepackage{amsfonts}
\usepackage{pbox}
\usepackage{makecell}
\makenomenclature
\usepackage{epstopdf}
\usepackage[pdftex]{graphicx}
\usepackage[export]{adjustbox}
\usepackage{ccaption}
\definecolor{green}{HTML}{66FF66}
\definecolor{myGreen}{HTML}{009900}
\usepackage{tcolorbox}

\usepackage{tabularx}
\usepackage{array}
\usepackage{colortbl}
\tcbuselibrary{skins}

\newcolumntype{Y}{>{\raggedleft\arraybackslash}X}

\tcbset{tab1/.style={fonttitle=\bfseries\large, fontupper=\normalsize\small, colupper=black!40!black,
colback=white!10!white,colframe=black!75!black,colbacktitle=black!40!black,
coltitle=black,center title,freelance,frame code={
\foreach \n in {north east,north west,south east,south west}
;}}}

\tcbset{tab2/.style={fonttitle=\bfseries, colupper=black!40!black,
colback=white!10!white,colframe=black!50!black,colbacktitle=black!40!black,
coltitle=black,center title}}

\tcbset{tab3/.style={enhanced,fonttitle=\bfseries,fontupper=\normalsize\small,colupper=black!40!black,
colback=white!10!white,colframe=black!50!black,colbacktitle=black!30!black,
coltitle=black,center title}}

\tcbset{tab4/.style={fonttitle=\bfseries\small,fontupper=\normalsize\small,colupper=black!10!black,
colback=white!10!white,colframe=black!10!black,colbacktitle=white!10!white,
coltitle=black,center title}}

\tcbset{tab5/.style={fonttitle=\bfseries\small,fontupper=\normalsize\small,colupper=black!10!black,
colback=white!10!white,colframe=black!10!black,colbacktitle=white!10!white,
coltitle=black,center title}}

\begin{document}
\title{Separation Framework: An Enabler for Cooperative and D2D Communication for Future 5G Networks}

\author{\IEEEauthorblockN{Hafiz A. Mustafa$^1$, Muhammad A. Imran$^1$, Muhammad Z. Shakir$^2$, \\Ali Imran$^3$, and Rahim Tafazolli$^1$}\\
\IEEEauthorblockA{Institute for Communication Systems (ICS), University of Surrey$^1$, Guildford, Surrey,UK\\
Electrical and Computer Engineering Dept., Texas A\&M University$^2$, Doha, Qatar\\
School of Electrical and Computer Engineering, University of Oklahoma$^3$, Tulsa, USA\\
Email: \{h.mustafa, m.imran, r.tafazolli\}@surrey.ac.uk$^1$,\\ muhammad.shakir@qatar.tamu.edu$^2$, ali.imran@ou.edu$^3$
}
}

\maketitle

\begin{abstract}
Soaring capacity and coverage demands dictate that future cellular networks need to soon migrate towards ultra-dense networks. However, network densification comes with a host of challenges that include compromised energy efficiency, complex interference management, cumbersome mobility management, burdensome signaling overheads and higher backhaul costs. Interestingly, most of the problems, that beleaguer network densification, stem from legacy networks’ one common feature i.e., tight coupling between the control and data planes regardless of their degree of heterogeneity and cell density. Consequently, in wake of 5G, control and data planes separation architecture (SARC) has recently been conceived as a promising paradigm that has potential to address most of aforementioned challenges. In this article, we review various proposals that have been presented in literature so far to enable SARC. More specifically, we analyze how and to what degree various SARC proposals address the four main challenges in network densification namely: energy efficiency, system level capacity maximization, interference management and mobility management. We then focus on two salient features of future cellular networks that have not yet been adapted in legacy networks at wide scale and thus remain a hallmark of 5G, i.e., coordinated multipoint (CoMP), and device-to-device (D2D) communications. After providing necessary background on CoMP and D2D, we analyze how SARC can particularly act as a major enabler for CoMP and D2D in context of 5G. This article thus serves as both a tutorial as well as an up to date survey on SARC, CoMP and D2D. Most importantly, the article provides an extensive outlook of challenges and opportunities that lie at the crossroads of these three mutually entangled emerging technologies.
\end{abstract}


\section{Introduction} \label{sec:introduction}
\IEEEPARstart{T}{raditional} cellular networks are designed with tight coupling of control and data planes. This architecture conforms to the main objective of ubiquitous coverage and spectrally efficient voice-oriented homogeneous services. The recent growth of data traffic overwhelmingly brought a paradigm shift from voice-traffic to data-traffic. Cisco made observations at internet service providers and predicted that the annual global Internet traffic will rise to 1.4 zettabyte by the year 2017 as compared to 528 exabyte (EB) in 2012 \citep{Cisco}. One of the contributors in this massive growth of Internet traffic is the proliferation of mobile devices and machine-to-machine (M2M) communication. Due to this growth, the capacity and coverage requirements exploded in recent years with worldwide mobile traffic forecast of more than 127 EB in the year 2020 \citep{UMTS_2011}. An increase of thousand-fold in wireless traffic is expected in 2020 as compared to 2010 figures \citep{6692781} with expected figure of 50 billion communication devices \citep{Ericsson_2011}. The explosive growth of mobile traffic is being handled by deploying tremendous amount of small cells resulting in heterogeneous network (HetNet)\nomenclature{HetNet}{Heterogeneous Network} \citep{6171992}.

The tight coupling of planes in conventional cellular networks leaves minimum control to consider networks' energy efficiency metric. This metric had a less concern previously due to less number of subscribers, rare data services, sparse deployments, and less awareness of green cellular communication. The green attribute of the cellular communication refers to reduction of unnecessary power consumption and its subsequent impact on the environment in the form of CO$_2$ emissions \citep{5677351,wu_green_2012,scott_matthews_planning_2010,5978416,6848019,6525595}. The green cellular communication can be realized by bringing energy-awareness in the design, in the devices \citep{4205092} and in the protocols of communication networks. Due to the network scaling and heterogeneity (large number of small cell deployments), this metric became prominent. In this regard, it has been estimated that the energy consumption by the information and communications technology (ICT) results in 2\% of global carbon emissions \citep{6100924}.

Small cell deployment is an agile, cost-effective, and energy efficient solution to meet coverage and capacity requirements. However, large number of deployments (e.g., prediction of 36.8 million small cell shipments by year 2016 according to ABI research \citep{ABI}), the energy efficiency gain due to small cells might be compromised. Moreover, it also poses operational expenditure (OPEX) challenges to the network operators. This heterogeneity has also imbalanced the provision of data services between macro and small cells resulting in severe interference/backhaul-limited communication. In order to overcome the threatening issues of power consumption, the awareness of energy consumption has already been realized and a number of energy conservation techniques/approaches have been investigated in the literature.

Another core issue, rising in future ultra-dense HetNet, is the interference management. The main limiting factor in achieving the optimum capacity is intra/inter-cell interference. Although intra-cell interference, in present cellular networks, has been eliminated by using orthogonal frequency division multiple access (OFDMA)\nomenclature{OFDMA}{Orthogonal Frequency Division Multiple Access} technology and radio resource management (RRM)\nomenclature{RRM}{Radio Resource Management}, provision of underlay co-existing networks (e.g., device-to-device (D2D), M2M), in future ultra-dense environment will again cause intra-cell interference along with existing inter-cell interference. Current interference management techniques mainly comprise mitigation, cancellation, and coordination. The first two techniques are best suited to a single cell environment, whereas for multicell scenarios, coordination techniques comprising inter-cell interference coordination (ICIC)\nomenclature{ICIC}{Inter-cell Interference Coordination}, enhanced ICIC (eICIC)\nomenclature{eICIC}{enhanced Inter-cell Interference Coordination}, coordinated beamforming (CB)\nomenclature{CB}{Coordinated Beamforming}, and coordinated multipoint (CoMP)\nomenclature{CoMP}{Coordinated Multipoint} are more promising to provide homogeneous quality of service with small infrastructural changes over the area \citep{4117538}. The ICIC techniques were introduced to mitigate inter-cell interference for cell-edge users. The main idea is to use either different set of resource blocks (RBs)\nomenclature{RB}{Resource Block} throughout the cell or partition RBs for cell-centre and cell-edge users. In another scheme of ICIC techniques, this RB partitioning can be coupled with different power levels (e.g., power boost for cell-edge users and low power for cell-centre users) to mitigate inter-cell interference. The ICIC techniques have been enhanced to eICIC for HetNet in $3^{\textnormal{rd}}$ generation partnership project (3GPP) \nomenclature{3GPP}{$3^{\textnormal{rd}}$ Generation Partnership Project} Rel-10. These techniques, unlike ICIC, consider both control and traffic channels either in time, frequency or power domains to mitigate inter-cell interference. The main idea of eICIC is based on almost blank subframe (ABS)\nomenclature{ABS}{Almost Blank Subframe}. These blank subframes are reserved for different purposes for macro tier and  small cell tiers. The macro tier mostly uses these subframes for control channels with low power, whereas small cell tiers use them for traffic channels to serve cell-edge users.

The CB and CoMP fall into the category of interference exploitation as compared to interference avoidance schemes (e.g., ICIC, eICIC). In such techniques, joint scheduling, transmission, and processing are carried out to exploit inter-cell interference and enhance cell-edge performance. In CB, the user equipment (UE)\nomenclature{UE}{User Equipment} is served by a single base station (BS)\nomenclature{BS}{Base Station}, however, interference is coordinated between cooperating BSs. To enhance the data rate of individual UE, it can cooperatively be served by a number of BSs in CoMP, however, this approach requires sharing data between cooperating BSs which results in huge backhaul capacity requirements. As compared to interference avoidance, the exploiting techniques (e.g., multicell cooperation (CB and CoMP)) have been identified as a key solution in long term evolution (LTE)\nomenclature{LTE}{Long Term Evolution} and LTE-Advanced (LTE-A)\nomenclature{LTE-A}{Long Term Evolution-Advanced} to improve the cell-edge performance, average data rate, and spectral efficiency by mitigating and exploiting inter-cell interference \citep{3gpp.36.819, marsch_coordinated_2011, 4657145}.

The green aspects of future 5G cellular networks require energy efficient communication which can be realized effectively by completely switching-off under-utilized BSs. However, the switch-off mechanism has severe limitations in current cellular architecture due to coverage holes. In order to avoid coverage holes, one of the candidate solution is the new cellular architecture where control and data planes are separated, i.e., decoupled or separation architecture, to provide ubiquitous coverage and more localized high-rate data services. Another potential advantage in this architecture is the flexible mobility management due to reduced handover signaling. In present architecture, the mobile user is handed over to nearby BS even if there is no active data session. Since, control plane is coupled with data plane, it is mandatory to handover in-active mobile terminals to ensure coverage. This results in handover signaling which is required for coverage but not for data services. On the other hand, the mobile user with no active data session in decoupled architecture can move freely without initiating handover due to ubiquitous coverage. Huge potential savings can be realized in this case, due to reduced handover signaling resulting in energy efficient communication.

In order to realize thousand-fold capacity enhancements in future cellular networks, much higher bandwidth is required. This higher bandwidth is available in millimeter wave (mm-Wave) spectrum. The higher frequency has poor propagation characteristics, however, the corresponding spot-beam coverage is more feasible for low-range high-rate data services. Therefore, coverage at lower frequencies (with good propagation characteristics) and high-rate data services (with limited coverage) requires decoupled architecture. Another aspect that severely limits the system capacity is the ultra-dense cellular environment in future networks (due to more granular tiers in the form of D2D, and M2M overlay/underlay communication). The underlay system offers higher system capacity but causes intra-cell interference and therefore, interference management becomes more complex in this case. For such an environment, cooperation and coordination is the promising solution for interference management in decoupled architecture. 

Keeping in view the above vision, we structure the article in three sections. The first section introduces separation framework and provides survey of existing literature on separation architecture. Since energy efficiency is the key enabler for separation framework, we provide extensive literature review of existing approaches that realize energy efficient communication in current cellular architecture. This is followed by highlighting future requirements of cellular networks from the perspective of system capacity, interference management, and mobility management. We highlight several shortcomings due to coupled planes and provide motivation for separation architecture. The shortcomings in current architecture and potential gains due to decoupling are tabulated at the end of first section. The second section provides a brief tutorial on cooperative communication including underlay D2D cooperation. This section serves as a background to discuss cooperation in separation architecture. The third section presents different scenarios where cooperative communication can be realized in separation framework by highlighting potential advantages and associated complexities. The article is organized as follows. We provide list of acronyms in Table \ref{Table:tabnom}. Section II  provides system performance reviews of traditional and separation architecture. In Section III, the general context of cooperative communication, clustering, and D2D communication for traditional cellular system has been presented. Section IV describes different perspectives to extend cooperative communication to the separation framework. Section V concludes the survey and highlights future research directions in this area.
\begin{table}[t]
\renewcommand{\arraystretch}{1.35}
    \centering
    \caption{List of Acronyms}\label{Table:tabnom}
    \vspace{2mm}
    \begin{tabular}{c||lc}
    	\hline
    	\textbf{Acronym}	&	\textbf{Definition}\\
    	\hline
    	3GPP	&	$3^{\textnormal{rd}}$ Generation Partnership Project\\
    	\hline
     	ABRB	&	Almost Blank Resource Block\\
    	\hline
	ABS	&	Almost Blank Subframe\\
    	\hline
	BBU	&	Base Band Unit\\
    	\hline
	BS	&	Base Station\\
    	\hline
	C-RAN&	Cloud Radio Access Network\\
    	\hline
	CARC	&	Conventional Architecture\\
    	\hline
	CB	&	Coordinated Beamforming\\
    	\hline
	cBS	&	Control BS\\
    	\hline
	CCU	&	CoMP Central Unit\\
    	\hline
	CoMP	&	Coordinated Multipoint\\
    	\hline
	CRC	&	Cyclic Redundancy Check\\
    	\hline
	CRS	&	Common Reference Signal\\
    	\hline
	CSI	&	Channel State Information\\
    	\hline
	CSI-RS&	Channel State Information Reference Signal\\
    	\hline
	CU	&	Central Unit\\
    	\hline
	dBS	&	Data BS\\
    	\hline
	eICIC	&	enhanced Inter-cell Interference Coordination\\
    	\hline
	eLA	&	enhanced Local Area\\
    	\hline
	eNB	&	evolved NodeB\\
    	\hline
	FDD	&	Frequency Division Duplex\\
    	\hline
	HARQ	&	Hybrid Automatic Repeat Request\\
    	\hline
	HetNet&	Heterogeneous Network\\
	\hline
	ICIC	&	Inter-cell Interference Coordination\\
    	\hline
	IMT-Advanced	&	International Mobile Telecommunications-Advanced\\
    	\hline
	ISM	&	Industrial, Scientific, and Medical\\
    	\hline
	JD	&	Joint Detection\\
    	\hline
	JT	&	Joint Transmission\\
    	\hline
	LTE	&	Long Term Evolution\\
    	\hline
	LTE-A	&	Long Term Evolution-Advanced\\
    	\hline
	MIMO	&	Multiple Input Multiple Output\\
    	\hline
	OFDMA&	Orthogonal Frequency Division Multiple Access\\
    	\hline
	PID	&	Physical Cell Identification\\
    	\hline
	PSS	&	Primary Synchronization Signal\\
    	\hline
	RAN	&	Radio Access Network\\
    	\hline
	RB	&	Resource Block\\
    	\hline
	RRH	&	Remote Radio Head\\
    	\hline
	RRM	&	Radio Resource Management\\
    	\hline
	RSRP	&	Reference Signal Received Power\\
    	\hline
	SARC	&	Separation Architecture\\
    	\hline
	SC-RAN&	SARC in C-RAN\\
    	\hline
	SON	&	Self-organizing Network\\
    	\hline
	SSS	&	Secondary Synchronization Signal\\
    	\hline
	TDD	&	Time Division Duplex\\
    	\hline
	UE	&	User Equipment\\
    	\hline
	UMTS & 	Universal Mobile Telecommunications System\\
	\hline
    \end{tabular}
\vspace*{-\baselineskip}
\end{table}
\section{Separation Framework: Performance Measures and Potential Gains}\label{sec:EE}
The current cellular networks comprise tightly coupled control and data planes in the same radio access network (RAN)\nomenclature{RAN}{Radio Access Network}. This architecture meets the main objective of ubiquitous coverage and spectral efficiency for voice services in homogeneous deployments. The massive growth of data traffic overwhelmingly dominated the voice traffic resulting into a paradigm shift from homogeneity to heterogeneity and voice services to data services. The traditional architecture (designed for homogeneous voice services) meets the current requirements of ubiquitous coverage and high spectral efficiency, however, it provides these services by overlooking signaling overheads, backhaul cost, and energy efficiency of the system. In order to enhance the coverage and capacity of current cellular systems, it is common practice to deploy small cells for peak-load scenarios at the cost of reduced energy efficiency, increased overhead signaling (e.g., in terms of frequent handovers) and increased backhaul requirements. In order to mitigate the rising concerns of power consumption, number of solutions, based on dynamic BS switching mechanism, are suggested to exploit the temporal and spatial variations in traffic load. However, the tight coupling of user and control planes restricts the flexibility and leaves less degree of freedom to optimize the system performance (discussed in subsequent discussions). To this end, the idea of control and data planes separation was proposed by the project beyond green cellular generation (BCG2) of GreenTouch consortium in Jan., 2011 \citep{greentouch}. Similar approaches have been suggested in study group of 3GPP on ``New Carrier Type". The Mobile and wireless communications Enablers for Twenty-twenty Information Society (METIS) \citep{METIS} aims to lay the foundation of 5G where control and data plane separation is being considered as a candidate system architecture. The green 5G mobile networks (5grEEn)  is focusing on green aspects of future 5G networks by considering separation of control and data planes. The joint European Union - Japan project Millimeter-Wave Evolution for Backhaul and Access (MiWEBA) is investigating the use of separated control and data planes for mm-Wave based small cells \citep{MiWEBA}.

\begin{figure*}[!htb]
\centering
    \subfigure[Conventional Architecture (CARC)]
    {
        \includegraphics[width = 3.25 in]{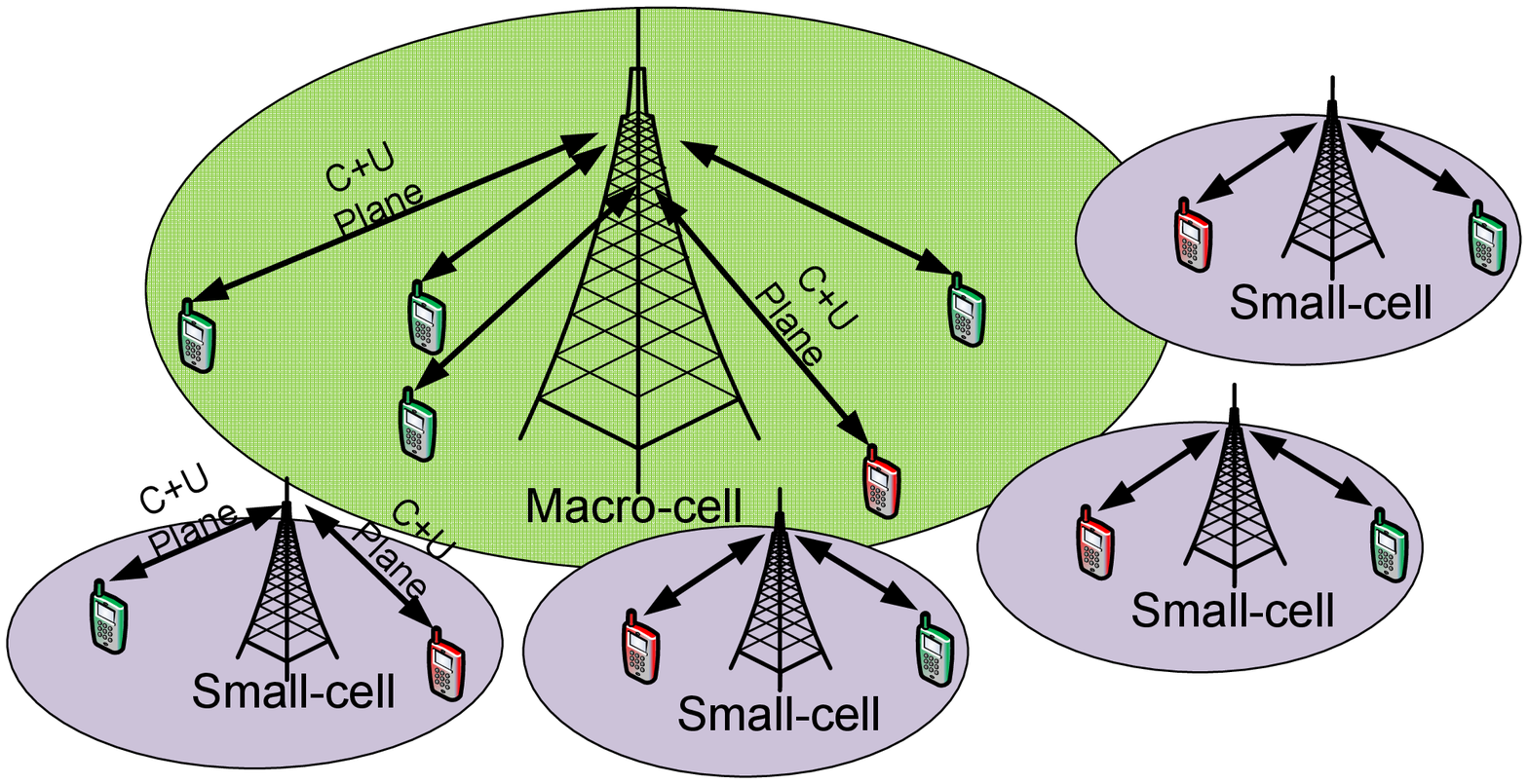}
        \label{Figure:carc}
    }
      \subfigure[Separation Architecture (SARC)]
    {
        \includegraphics[width = 3.25 in]{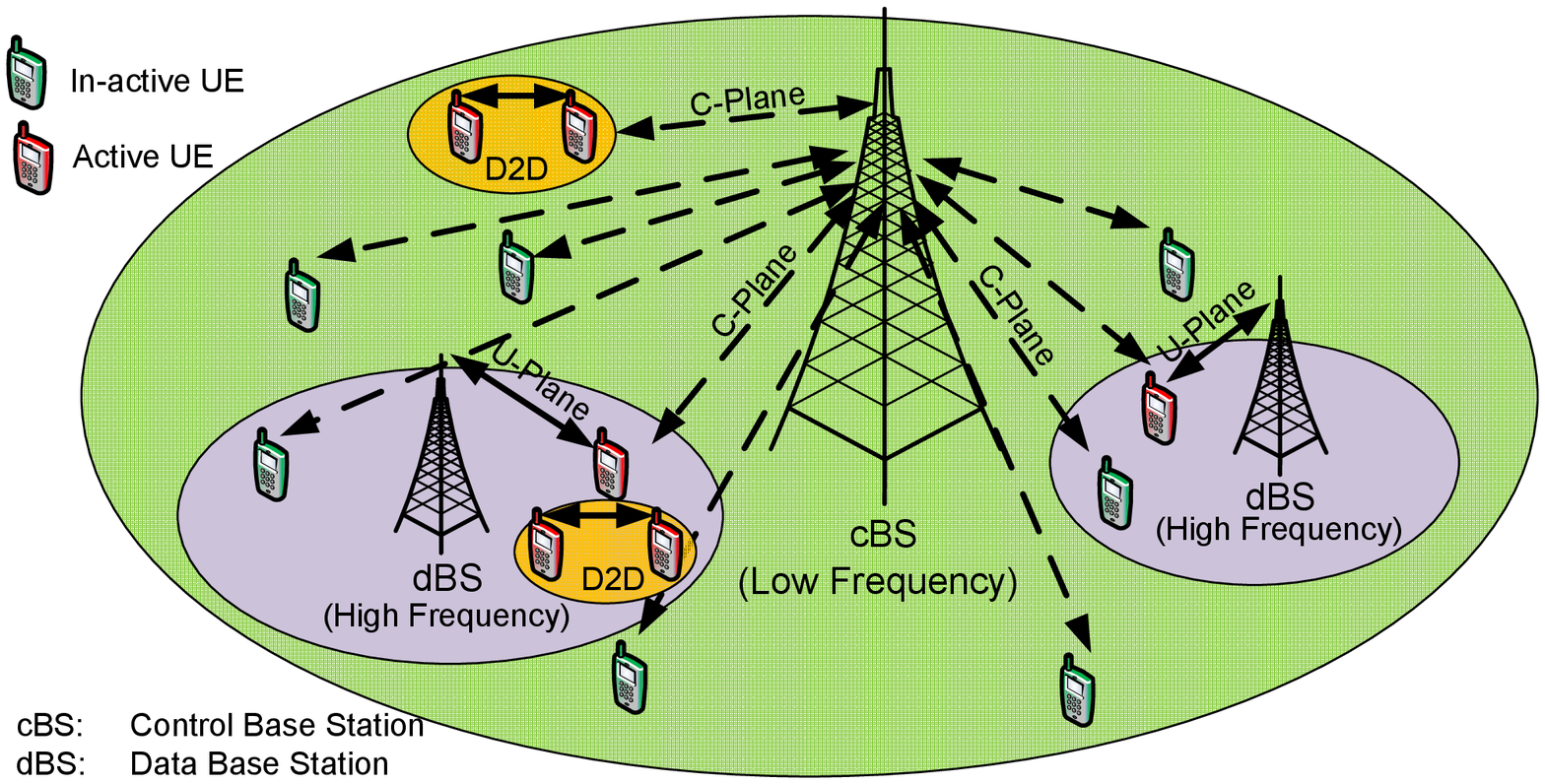}
        \label{Figure:sarc}
    }
 \caption{Conventional and control/data plane separation architecture.}
   \vspace{-3mm}
\label{Figure:carc_sarc}
\end{figure*}

In order to highlight potential gains due to decoupling of control and data planes, we present conventional architecture (CARC)\nomenclature{CARC}{Conventional Architecture} and futuristic separation architecture (SARC)\nomenclature{SARC}{Separation Architecture} in Fig. \ref{Figure:carc_sarc}.
As shown in Fig. \ref{Figure:carc}, CARC is a conventional HetNet (comprises macrocell and large number of small cells) where coverage and data services are simultaneously provided at same frequency either by macro or small cell on coupled control and data planes. The advantage of this approach is ubiquitous coverage, however, the serving cell cannot sleep and it has to provide coverage even at low load conditions resulting in under-utilization of resources. The mobile users, irrespective of active or in-active sessions, are always covered by dedicated channels (ubiquitous coverage). However, it results in under-utilization of data plane (since it is coupled with control plane). In Fig. \ref{Figure:sarc}, SARC is a hierarchical HetNet comprising conventional HetNet and an additional tier of D2D/M2M communication, where control and data planes are decoupled. In such an architecture, the ubiquitous coverage and low-rate data services\footnote{The control BS has ubiquitous coverage over a large area as compared to small cell coverage area. Hence, it is more feasible to provide data services to high mobility users by cBS to avoid signaling overhead and frequent handovers in small cells.} are provided by control BS (cBS)\nomenclature{cBS}{Control BS} at lower frequency bands with good channel characteristics. The data services are provided on demand at higher frequency bands by short range high-rate data BSs (dBSs)\nomenclature{dBS}{Data BS}. The advantages of this architecture are ubiquitous coverage (by decoupled control plane for active or in-active users), small cell sleeping possibility without coverage holes, temporal and spatial traffic adaptation, and high-rate data services for active users without compromising the energy efficiency of the system. The reader is referred to \citep{6468982} for feasibility study of detached cells from the perspective of reliability and energy savings.

The control plane is responsible for system configuration and management. It provides system information, synchronization, and reference signals etc. The system information is broadcast and it mainly comprises the information required to join the network. The synchronization information includes frame timings as well as symbol level timings. The reference signals are used to know channel state which is indispensable for scheduling and resource allocation. In contrast to this, data plane is responsible to provide the requested contents along with some acknowledging mechanism (e.g., hybrid automatic repeat request (HARQ)\nomenclature{HARQ}{Hybrid Automatic Repeat Request}). In order to give insights into information exchanges in both the planes, we provide a case study of LTE/LTE-A networks in Table \ref{Table:tab2}.
\begin{table*}[!htb]
    \begin{center}
    \caption{Control data plane information exchange in LTE/LTE-A.}\label{Table:tab2}
    \vspace{2mm}
\begin{tcolorbox}[tab2,tabularx={p{1.2in}||X|c||c}]
\bf{Signals} & \bf{Information Exchange} & \bf{Direction} & \bf{Plane}  \\ \hline

Physical random access channel (PRACH) & Initial synchronization with eNode B (eNB). & \multirow{7}{*}{Uplink}& \multirow{6}{*}{Control}\\ \cline{1-2}

Reference Signals (RS) & Demodulation RS (DRS) - Channel estimation for coherent demodulation, Sounding RS (SRS) - Channel quality estimation over a span of bandwidth. &  &  \\ \cline{1-2}

Physical uplink control channel (PUCCH) & (HARQ ACK/NACK)*, channel quality indicator (CQI), precoding matrix indicator (PMI), rank indicator (RI), scheduling requests. & & \\ \cline{1-2} \cline{4-4}

Physical uplink shared channel (PUSCH) & User uplink data. & & Data \\ \hline

Synchronization & Primary and secondary synchronization (PSS, SSS) for cell identity and frame timing. & \multirow{8}{*}{Downlink} & \multirow{7}{*}{Control} \\ \cline{1-2}

Reference Signals (RS) & Channel state information (CSI-RS), demodulation (DM-RS), cell-specific (CRS), positioning (PRS). &  & \\ \cline{1-2}

Control Indicators & Physical control format indicator channel (PCFICH) to indicate size of PDCCH, physical HARQ indicator channel (PHICH) to ACK/NACK user data on PUSCH. & & \\ \cline{1-2}

Multicast/Broadcast & Physical broadcast channel (PBCH) carrying master information block (MIB), multicast/broadcast single frequency network (MBSFN), multicast channel (PMCH). & & \\ \cline{1-2} \cline{4-4}
 
Physical downlink shared channel (PDSCH) & User multiplexed data. & & Data \\ \hline

\multicolumn{4}{l}{\multirow{1}{*}{* HARQ is sent either as a feedback message on control channel or piggybacking feedback on user's data plane.}} \\ [0.25ex]
\end{tcolorbox}
\vspace{-6mm}
  \end{center}
\end{table*}

The SARC for HetNet offers many potential gains such as energy efficiency, capacity enhancement, reduced overhead signaling, flexible interference and mobility management. Control signaling is provided by cBS, however, certain types of control signaling cannot be fully decoupled. For example, frame/symbol level synchronization and channel state information (CSI)\nomenclature{CSI}{Channel State Information} is required in both planes.

The separation of planes for future cellular networks has been realized very recently. To this end, the control and data plane separation has been suggested in \citep{6152217, 6515050}, where the provision of coverage has been provided by a long range low rate control evolved Node B (eNB)\nomenclature{eNB}{evolved NodeB}. The data services, on the other hand, are provided by dedicated data eNBs. In \citep{6152217}, it is proposed that signaling will provide wider coverage to all UEs regardless of active or in-active data session under data eNB. Such network-wide adaptation provides flexibility to power down certain BSs when no data transmission is needed. In simple strategy of powering down the dBSs, neither control signaling (e.g., synchronization, reference signals, system information etc) nor associated backhaul to the access network is required; no data services are requested by UEs, only coverage is required which is ubiquitously provided by the cBS. The powering down strategy can, therefore, save approximately 80\% of RAN power per BS switch-off \citep{6152217, 6056691,4448824} besides power savings due to backhaul communication links. Therefore, separation of planes promises tremendous increase in energy efficiency, reduced overhead signaling, and relaxed backhaul requirements. In \citep{6152217}, the energy efficiency gain has been emphasized by considering system level approach where under-utilized BSs are realized in sleep mode. In this study, no expected gains in energy efficiency are highlighted. Certain technical challenges including context awareness, resource management, and radio technologies for the signaling network are highlighted without proposing any design guidelines for the separation architecture.

The design of the signaling network in SARC is more challenging as compared to conventional approach. In CARC, the BSs usually do not sleep due to the possibility of coverage holes. Therefore, all BSs are active and no wake-up signaling is required. The handover procedure is usually UE driven based on reference signal received power (RSRP)\nomenclature{RSRP}{Reference Signal Received Power} values. In contrast, data services in SARC, in case of sleeping dBS, can be ensured by (i) optimal dBS selection from sleeping dBSs, and (ii) initiation of wake-up mechanisms. The optimal dBS selection can be quite challenging since cBS has no instantaneous knowledge of channel conditions. This results in more complex signaling procedures as compared to CARC. The new design is required to be robust and energy efficient. Use of low frequencies provides better propagation and obstacle penetration. Moreover, mobility management is flexible in HetNet using SARC architecture. This is because, control plane handover is rarely required since the coverage area of cBS is large as compared to the coverage area of BSs in conventional system. The data plane handover is only required in case of active data requests and in case of in-active users, none of the handovers (control plane or data plane) are required. This has been discussed in more details in Sec. \ref{MM:ho_proc}.

In \citep{6515050}, a two-layer network functionality separation scheme, targeting low control signaling overhead and flexible network reconfiguration for future green networks has been proposed. A frame structure level detail has been proposed in which network functionality including synchronization, system information broadcast, paging, and multicast (synchronization, pilot, frame control, and system/paging/multicast information bearer signals) is incorporated in control network layer (CNL). Whereas, the network functionality of synchronization and unicast (synchronization, pilot, frame control, and unicast information bearer signals) is incorporated in data network layer (DNL). In this study, the main focus is given on advantages of low control signaling overhead. The network area power consumption has been plotted for two architectures showing significant potential gain for separation architecture leading towards future energy efficient green mobile networks. Unlike \citep{6152217}, the authors in \citep{6515050} proposed abstract level network design for control and data planes separation. The categorization of different wireless signals and their mapping relationship with physical channels are presented. However, the challenges highlighted in \citep{6152217} are not discussed in \citep{6515050}. The study also lacks in addressing interference management issues, backhaul requirement, realization of underlay networks (e.g., D2D), mobility management and corresponding handover procedures in separation architecture.

The important focus areas for energy efficient 5G mobile network are highlighted in \citep{6673363}. These areas include system architecture with decoupled control and data planes, ultra-dense HetNet deployment, radio transmission using multiple input multiple output (MIMO)\nomenclature{MIMO}{Multiple Input Multiple Output} configuration and energy efficient backhaul. The transmission planes are categorized into data, control, and management planes. It is emphasized that if these planes are decoupled from each other then independent scaling is possible at most energy efficient locations. Furthermore, the logical separation of control and data planes can provide most efficient discontinuous transmission/reception (DTX/DRX) functionality to save energy in idle modes. Similar to \citep{6152217}, the authors in \citep{6673363} highlighted the requirements and technical challenges to realize future green 5G mobile network. However, the system architecture and radio transmissions design guidelines are not outlined in details as in \citep{6515050}. The solutions to these important areas are considered as deliverables of 5GrEEn.

In \citep{zhao2013software}, hyper-cellular network is introduced as decoupled control and traffic network to realize energy efficient operation of BS. In such a network, data cells are flexible to adapt traffic variations and network dynamics while control cells can flexibly and globally be optimized. The hyper-cellular network is considered as a novel architecture for future mobile communication systems. The approach realizes control and data planes separation using open source radio peripherals and legacy global system for mobile (GSM) network. In this testing, signaling BS provided coverage whereas data BS ensured phone call connectivity. A very promising formulation has been setup by using open base transceiver station (OpenBTS), universal software radio peripheral (USRP) front end, wide bandwidth transceiver (WBX) daughter board, and dell PCs. This formulation provides an insight into real-time practical setup for prototype testing. However, system improvements are not shown in this paper. Moreover, none of the performance metrics (energy efficiency, backhaul relaxation, and throughput) have been analyzed and validated for this simple and basic approach.
\begin{table*}[!htb]
\makeatletter
\newcommand*{\compress}{\@minipagetrue}
\makeatother
\renewcommand{\arraystretch}{1.5}
\caption{Summary of approaches for control and data planes separation.}\label{Table:Appr_CDplane}
\vspace{2mm}
\begin{tcolorbox}[tab1,tabularx={>{\raggedright\arraybackslash}p{1.1in}||>{\raggedright\arraybackslash}p{1in}|X|>{\raggedright\arraybackslash}p{1.45in}}]
\textbf{Project/Paper/Ref.}			&\textbf{Aim}					&\textbf{Working/Highlights} 														& \textbf{Impacts/Conclusion}					\\ \hline\hline
``\textit{Looking beyond green cellular networks}" \citep{6152217}			
					&  {To switch-off BSs flexibly in case of no data transmission}\vspace*{-\baselineskip}
												& \compress \begin{itemize}[leftmargin=1.25em]
													\renewcommand{\labelitemi}{$\Rightarrow$}
													\item Coverage $\rightarrow$ by long-range low-rate control eNB 
													\item Data Service $\rightarrow$ by short-range high-rate data eNBs.
													\item Ubiquitous coverage by signaling plane.
													\vspace*{-\baselineskip}
												\end{itemize}																		&\compress\begin{itemize}[leftmargin=0.75em]
																																	\item Selection and activation of BS is not a difficult task compared to optimizing the decision process.
																																	\vspace*{-\baselineskip}
																																\end{itemize} 						\\ \hline
``\textit{On Functionality Separation for Green Mobile Networks: Concept Study over LTE}" \citep{6515050}
					& To reduce control signaling overhead \& realize flexible network reconfiguration
												&\compress\begin{itemize}[leftmargin=1.25em]
													\renewcommand{\labelitemi}{$\Rightarrow$}
													\item Separation scheme based on two-layer network functionality: CNL/DNL
													\item CNL $\rightarrow$ multicast information bearer signals
													\item DNL $\rightarrow$ unicast information bearer signals
													\vspace*{-\baselineskip}
												\end{itemize}														
																																& \compress\begin{itemize}[leftmargin=0.75em]
																																	\item Rare Handover in CNL 
																																	\item Call re-establishment is not required 
																																	\item HO signaling is reduced significantly
																																	\vspace*{-\baselineskip}
																																\end{itemize}						\\ \hline 
``\textit{5GrEEn: Towards Green 5G mobile networks}" \citep{6673363}
					& To provide general outlook on system architecture for energy efficient 5G network
												&\compress\begin{itemize}[leftmargin=1.25em]
													\renewcommand{\labelitemi}{$\Rightarrow$}
													\item Ultra-dense HetNet deployment
													\item Radio transmission using MIMO configuration 
													\item Energy efficient backhaul
													\item Transmission planes: data, control, and management
													\vspace*{-\baselineskip}
												\end{itemize}														
																																& \compress\begin{itemize}[leftmargin=0.75em]
																																	\item Separation of control and data plane provides most effective DTX/DRX functionality to save energy in idle modes
																																	\vspace*{-\baselineskip}
																																\end{itemize}						\\ \hline 
``\textit{Software defined radio implementation of signaling splitting in hyper-cellular network}" \citep{zhao2013software}
					& Energy efficient operation of BS
												&\compress\begin{itemize}[leftmargin=1.25em]
													\renewcommand{\labelitemi}{$\Rightarrow$}
													\item Hyper cellular network: Decoupled signaling and data services
													\item Handset is provided coverage by signaling BS 
													\item Phone calls are connected with the help of data BS
													\vspace*{-\baselineskip}
												\end{itemize}														
																																& \compress\begin{itemize}[leftmargin=0.75em]
																																	\item Provides an insight into real-time practical setup for prototype testing
																																	\vspace*{-\baselineskip}
																																\end{itemize}						\\ \hline
``\textit{FP7 project CROWD}" \citep{6702534}
					& Energy optimized connectivity management
												&\compress\begin{itemize}[leftmargin=1.25em]
													\renewcommand{\labelitemi}{$\Rightarrow$}
													\item SDN based MAC control and mobility management
													\vspace*{-\baselineskip}
												\end{itemize}														
																																& \compress\begin{itemize}[leftmargin=0.75em]
																																	\item Complements huge deployments of cellular nodes
																																	\vspace*{-\baselineskip}
																																\end{itemize}						\\ \hline
``\textit{Dual connectivity in LTE HetNets with split control- and user-plane}" \citep{6825019}
					& Dual connectivity and use of CSI-RSs for CSI measurements
												&\compress\begin{itemize}[leftmargin=1.25em]
													\renewcommand{\labelitemi}{$\Rightarrow$}
													\item Different MA small cells are considered as different antennae of MIMO/CoMP array
													\vspace*{-\baselineskip}
												\end{itemize}														
																																& \compress\begin{itemize}[leftmargin=0.75em]
																																	\item CSI-RS is also used to estimate the downlink path loss for uplink power control
																																	\vspace*{-\baselineskip}
																																\end{itemize}						\\ \hline
``\textit{A novel architecture for LTE-B: C-plane/U-plane split and Phantom Cell concept}" \citep{6477646,6554746}
					& To provide high data rate to UE through spatial reuse of spectrum
												&\compress\begin{itemize}[leftmargin=1.25em]
													\renewcommand{\labelitemi}{$\Rightarrow$}
													\item Phantom Cell architecture: high frequency band solution with decoupled control/data plane
													\item Macro cell controls the small cells for connection establishment 
													\item Small cells use high frequency bands to provide high-rate data coverage
													\vspace*{-\baselineskip}
												\end{itemize}														
																																& \compress\begin{itemize}[leftmargin=0.75em]
																																	\item Outperforms conventional small cell architecture in both spectral and energy efficiency metrics
																																	\vspace*{-\baselineskip}
																																\end{itemize}						\\  \hline\hline
\end{tcolorbox}
\vspace{-4mm}
\end{table*}

The control and data planes separation concept has been presented from the perspective of energy optimized connectivity management in seventh framework programme (FP7) CROWD \citep{6702534}. To this end, software defined networking (SDN) based medium access control (MAC) and mobility management has been proposed to complement huge deployments of cellular nodes. Two key challenges, interference and mobility management, are considered for next generation dense wireless mobile networks. The functional architecture has been proposed and several key control applications are identified. More focus is given on mobility management and an SDN-based distributed mobility management (DMM) approach has been suggested. The control applications for interference management range from existing multi-tier scheduling scheme (e.g., eICIC) to LTE access selection schemes. The radio transmission aspects and backhaul limitations have not been outlined in any of the control applications identified in this study.

In \citep{6825019}, the authors measure CSI by using the concept of dual connectivity (using macrocell assisted small cells) and proposing the use of CSI reference signals (CSI-RSs)\nomenclature{CSI-RS}{Channel State Information Reference Signal} instead of common reference signal (CRS) \nomenclature{CRS}{Common Reference Signal}. Since, CSI-RSs are traditionally used by UEs to differentiate between different antennas of a MIMO system, therefore in the proposed network layout, different macrocell assisted small cells are considered as different antennae of MIMO/CoMP array. This strategy results in energy efficient operation (by reducing number of CRS) and provides network-triggered handover (unlike UE-triggered handover in CARC) to realize flexible and enhanced mobility management. Due to the absence of CRS for macrocell assisted small cells, the authors proposed to use CSI-RSs to estimate the downlink path loss for uplink power control. Similar to the previous approaches, the authors in \citep{6825019} focuses only on reducing control signaling to realize energy efficient operation without emphasizing context awareness, radio frame structure, backhaul issues, and interference management.

The 3GPP is presently standardizing enhanced local area (eLA)\nomenclature{eLA}{enhanced Local Area} small cell HetNet (LTE Rel-12) to provide high data rate to UEs through spatial reuse of the spectrum. In \citep{6477646}, a particular eLA architecture called Phantom Cell is proposed by NTT DOCOMO. This architecture is based on control and data planes separation; suggested as a novel architecture for LTE-B. The approach in \citep{6477646} suggests deployment of massive small cells by leveraging high frequency reuse under the coverage of macrocell to achieve high capacity, seamless mobility, and scalability. The two tier configuration is realized as a master-slave configuration where macrocell controls the small cells dynamically for connection establishment and small cells use high frequency bands to provide high-rate data coverage. This high frequency band solution with decoupled control and data planes, where small cells do not transmit cell-specific reference signals, is introduced as Phantom Cell architecture. In order to evaluate the energy efficiency performance of the Phantom Cell architecture, the stochastic geometry is used to compare the results with the conventional frequency division duplex (FDD)\nomenclature{FDD}{Frequency Division Duplex} based LTE picocell deployment in \citep{6554746}. The numerical results indicate that the Phantom Cell architecture outperforms conventional small cell architecture in both spectral and energy efficiency metrics. The authors in \citep{6477646} provide preliminary results for capacity enhancements in separation architecture without considering energy efficiency aspects, whereas \citep{6554746} provides more rigorous analysis for both spectral and energy efficiency of separation architecture. Some interesting conclusions are made about higher spectral efficiency and higher energy efficiency, however, both these studies focused on spectral and energy efficiency metric and did not include other aspects such as context awareness, signaling network, and functional description of the separation architecture. The reader is referred to \citep{7067574, 6848637} for Phantom cell operation at super high and extremely high frequency and related technical issues such as larger path loss in small cell, human body shadowing, massive MIMO architecture, and precoding algorithms to achieve super high data rates. The comparative summary of different approaches for control and data planes separation is presented in Table \ref{Table:Appr_CDplane}.

In the following subsections, we provide motivation for control and data planes separation architecture. In this context, we consider several key performance measures and analyze them in existing architecture. We provide survey of existing approaches, highlight the shortcomings and discuss these measures from the perspective of SARC architecture.
\subsection{Energy Efficiency}\label{ee}
The energy efficiency of RAN mainly depends on power consumption of BS. According to energy aware radio and network technologies (EARTH) project \citep{6056691}, the BS power consumption model comprises power consumed by radio frequency chain (especially power amplifier), signal processing units, and supply units (mains supply, DC-DC, and active cooling) as follows:
\begin{align}
\mathcal{P}_{BS} \,\,\, \alpha \,\,\, (\mathcal{P}_{RFC}, \mathcal{P}_{SPU}, \mathcal{P}_{SU}),	\IEEEnonumber
\end{align}

In order to ensure energy efficient communication, one simple strategy can be adopted where under-utilized BS, in case of low traffic conditions, should go to sleep mode (hence reducing power consumption $\mathcal{P}_{RFC}$ and $\mathcal{P}_{SU}$). This situation, however, causes coverage holes due to tight coupling of control and data planes unlike futuristic architecture where coverage and data services will be decoupled to provide ubiquitous coverage and on-demand data services.

The power consumption had not been a problem in past due to homogeneous networks and sparse deployments. Therefore, energy efficiency metric had not been considered while designing such cellular networks. Due to technology scaling and proliferation of large number of smart devices, the capacity demands increased tremendously with more energy consumption worldwide. This huge increase in capacity was predicted by wireless world research forum (WWRF) more than a decade ago. The key technological vision from WWRF expected around 7 trillion wireless devices serving 7 billion people by 2017. Moreover, it was predicted that approximately 80-95\% subscribers will be mobile broadband users \citep{wwrf2009,tafazolli2006technologies}. The huge increase in number of subscribers motivated the network operators to deploy small cells in order to quickly meet the customer needs. According to ABI research, by 2016, small cells will cover up to 25\% of all mobile traffic and small cells shipments (both indoor and outdoor) will likely to reach 36.8 million units worth \$20.4 billion. It further predicts that, outdoor small cell units alone will reach over 3.5 million units by 2018 \citep{Huawei,FierceMobileIT,SCFORUM,ABI}. The coverage and capacity requirements of subscribers can be met by deploying increased number of small cells, however, the associated power consumption will increase significantly in future. 

In order to reduce the power consumption of under-utilized BSs and ensure energy efficient communication in existing HetNet, different techniques are reported in literature such as dynamic BS switch-off, cell range expansion etc. These techniques provide substantial gain in power saving, however, they come with the inherent problem of coverage holes (in case of BS switch-off), increased interference (due to increased transmit power in cell range expansion techniques), and huge backhaul requirements. To address these problems and ensure energy efficient communication, a paradigm change in control and data planes coupling has been suggested in literature and research community. This approach not only provides ubiquitous coverage and reduced transmit power but also reduces control signaling associated with each BS.

The current cellular systems are designed for worst case ubiquitous coverage scenarios. In such a design, the BS needs to be active even for few subscribers. This goal can be justified in remote sparsely populated areas covered by few BSs where the spatio-temporal variations of traffic patterns follow a near constant trend. However, in urban areas, the BS deployment is dense and traffic variations are more abrupt. In such dense deployments, the coverage goal is achieved at the cost of increased power consumption and reduced energy efficiency of the system. The most power expensive element of RAN is BS, consuming around 80\% of overall power \citep{6152217,4448824}. In full-load conditions, the power consumption of BS is justified, however, in low load conditions, BS is still consuming most of the power to provide coverage. Moreover, in design of cellular systems, the short-term and long-term traffic variations (e.g., temporal effects on traffic loads due to day/night times and spatial effect due to weekends/weekdays) are not considered due to which the existing cellular networks cannot be fully optimized from this perspective.
\subsubsection{Evaluation Framework}\label{sec:efw}
In order to quantify the power consumption of wireless networks, the EARTH project provides a holistic energy efficiency evaluation framework (E3F). This framework provides power consumption breakdown of each entity of RAN. A BS power model has been proposed that maps the radio frequency (RF) output power (radiated from the antenna) to the total supply power consumption of BS. The power consumption for macro, micro, pico, and femto cells are compared. The traffic models (short-term and long-term) are investigated to emphasize the energy saving potentials. The deployment areas of Europe are segregated into dense urban, urban, suburban, rural and sparse. The traffic variations for a single day are depicted to give an insight into the energy efficiency evaluation of the wireless cellular network. Number of key findings are presented as follows:
\begin{itemize}
  \item On average, the vast majority of the resources are idle in wireless networks.
  \item The supply power scales linearly to the number of transmit/receive chains.
  \item The RF output power and power consumption of BS are nearly linear.
  \item For macro BSs, the consumption of power amplifier (PA) scales with BS load.
  \item For micro BSs, the PA scaling is present to a lesser extent, whereas for pico/femto BSs, this scaling is negligible.
\end{itemize}

It has been mentioned that DC power consumption of a typical 3-sector site at zero load is still 50\% of the peak power \citep{5722322}. The conventional model without power supply and active cooling/air conditioning can be 400W lower than the total power consumption of a site \citep{5621969}. In \citep{6600717}, a parameterized linear power model is proposed to encompass the two general power saving techniques that are based on either design change or operating procedures. The former is based on changing the layout of the network (e.g., by introducing HetNet) whereas the latter is more attractive for existing architecture. This approach saves energy by reducing transmission power, adapting transmission bandwidth, deactivating unused antennas, and incorporating BS sleep modes.

The model presented in \citep{6600717} is the simpler parameterized model of \citep{6056691}. The authors did not discuss the implications of coverage holes due to sleep mode operation. 
\subsubsection{BS Switch-off}\label{sec:BS_sw_off}
Power consumption of cellular systems has been addressed from two perspectives. The first one motivates the use of low-powered components in cellular networks and hence focuses on reducing the energy consumption at local scope. The second perspective takes the holistic approach of network design, planning, and management phases to conserve the energy of the overall cellular network. In both cases, the most power expensive element in access network is BS. A lot of research has been carried out to propose switch-off mechanisms for BSs. In \citep{6629715}, BS switch-off has been proposed by quantifying the reduction in activity probability for cooperative scenario. It has been shown that for a fixed distance between BSs, the expected number of enabled BSs reduces up to 11\% depending on the user density. By changing the distance  between cooperative BSs to an optimal value, an additional 39\% reduction in activity probability can be achieved which results corresponding reduction in power consumption per unit area. The proposed analysis assumed perfect hexagonal grid which is non-realistic in practical BS deployments. Moreover, finding the optimal distance and changing the BS deployment is practically infeasible and very hard to realize. The authors in \citep{7037473} suggest probabilistic data BS sleeping mechanism in separation architecture. The formulated problem jointly optimizes the sleeping probability and spectrum resource allocation to minimize the overall power consumption, however, this study does not consider mobility of the users and their impact on cell sleeping probability.

In \citep{5208045, 5300273, 5683654}, traffic profile based BS switching has been proposed to save energy. The cell switch-off has been suggested for cellular access networks \citep{5208045} and universal mobile telecommunications system (UMTS) \nomenclature{UMTS}{Universal Mobile Telecommunications System} access networks \citep{5300273} with the assumption that the radio coverage will be provided by neighboring cells by increasing transmit power. The smaller number of BSs for long-term switch-off vs. larger number of BSs for short-term switch-off have been investigated. However, these studies considered ideal networks (hexagonal and manhattan models) and introduced the energy saving by dynamic switching algorithms. Though both of the approaches \citep{5208045, 5300273} target to reduce power consumption, this strategy can cause severe interference to the neighboring active BSs due to increased transmit power. This can be ideal for the scenario where all neighboring BSs need switch-off which is not practical. The approach in \citep{5683654} considered first and second order statistics of traffic profile to propose dynamic switching strategy. The users are handed over to the neighboring cells before the reference cell can be switched off. The statistics based switching strategy can save energy, however, it is suitable for near-constant traffic pattern (e.g., night times). In case of slowly varying traffic profile, an instantaneous switching strategy is more promising which can flexibly be realized in SARC. A simple approach may consider traffic profile and provide data service either by cBS (in case of low-data rates) or by the near-by dBS (in case of high-data rates). In case of sleeping dBS, the cBS (having global coverage) may initiate wake-up mechanism which can be reactive or pro-active by predicting user mobility patterns. Since, no transmit power of dBS is increased, therefore an energy efficient communication, without increasing interference, can be achieved in SARC as compared to approaches proposed for CARC.

The macro BSs provide bigger coverage with high transmit power as compared to small cells. In order to conserve energy, the capacity enhancements are carried out by deploying large number of low-powered small cells. This brings heterogeneity in the network. For such networks, an area power consumption metric has been investigated in \citep{5360741},\citep{rost201011}, to quantify energy savings. The small cell deployment offers substantial power savings, however, this strategy scales poorly with number of small cells envisioned for future ultra-dense cellular environment. The scaling of small cells can be compensated by dynamic BS switch-off mechanisms which can be realized in separation architecture without producing coverage holes. The approaches in \citep{5360741, rost201011} considered mixed deployment scenarios by considering macro and micro cells at fixed positions. This strategy is suitable for new deployments but it is not applicable to existing deployments of small cells. Assuming perfect hexagonal grid is a theoretical interest. These studies also lack in presenting realistic operating algorithm where area power consumption scales with any change in deployment e.g., due to network scaling or BS failure.

The BS switching-on/off based energy saving (SWES) algorithm has been proposed in \citep{6489498} to exploit the temporal and spatial variation in the network traffic profile. The algorithm works in a distributed manner with reduced computational complexity. A notion of network-impact has been introduced that ensures minimal effects on neighboring BSs by turning off BSs gradually (one by one). In order to reduce overheads over the air and backhaul, three other heuristic versions of SWES are proposed that take network-impact as decision metric. The authors claim around 50-80\% potential savings for real traffic profile of metropolitan urban area. Several extensions of this research are proposed as follows:
\begin{itemize}
  \item To consider more realistic BS power consumption model.
  \item To consider HetNet, consisting of different types of BSs, such as macro, micro, femto BSs and even WiFi APs.
  \item To develop a dynamic BS switching algorithm that considers downlink and uplink traffics jointly.
\end{itemize}

Besides these extensions, the authors did not consider quality-of-service (QoS) requirements of the handed-over users. For example, in homogeneous deployments with large coverage area, cell-centre users have certain QoS requirements. In case, the serving BS of these users has to be switched-off, the neighboring BSs cannot guarantee same QoS without increasing the transmit power which results in inter-cell interference. This situation can be avoided in HetNet where neighboring small cells can cover handed-over users with moderate increase in transmit power, however, SWES techniques are proposed for only homogeneous deployments.

The theoretical framework for BS energy saving is presented in \citep{5992823}. It encompasses dynamic BS operation, and related problem of user association together. The problem is formulated as total cost minimization that allows for a flexible trade-off between flow-level performance (e.g., file transfer delay) and energy consumption. For user association problem, an optimal energy-efficient user association policy has been proposed, whereas for BS operation problem (i.e., BS switching-on/off), a simple greedy-on/off algorithm, based on mathematical background of sub-modularity maximization problem, is proposed. A number of heuristic algorithms, based on the distances between BSs or the utilization of BSs that do not impose any additional signaling overhead, are also proposed. The numerical results show 70-80\% reduction in total energy consumption while depending on the arrival rate of traffic and its spatial distribution as well as the density of BS deployment. Unlike \citep{6489498}, the theoretical framework in \citep{5992823} considers HetNet, however, to ensure mathematical tractability, no fast fading is considered and inter-cell interference is assumed as Gaussian-like noise which restricts practical realization of the proposed technique.
Since an under-utilized BS consumes nearly the same power as a fully loaded BS \citep{6056691}, the logical solution to this problem is to switch off idle BSs while providing the same coverage and quality of the service. However, switching-off BSs will create coverage holes as the signaling and data services are provided by the same BS. A number of different techniques are proposed in literature to solve this problem. A paradigm shift in control and data planes coupling has been suggested in \citep{6152217} where the coverage is provided by long-range BSs and high-rate data services are provided by small cell BSs. Hence, these short-range BSs can be activated/deactivated according to user demands without creating coverage holes.
\subsubsection{Renewable Energy Resources}\label{sec:re-new res}
The energy efficiency of cellular systems has also been addressed using renewable energy sources. The cellular networks are scaled according to developed environment (e.g., urban, sub-urban, rural) and network traffic, however, the rural areas usually dominate on a country-wide coverage \citep{6056691}. In developing countries, many remote locales do not have access to national electricity grid. To provide coverage in these areas, usually diesel is used as an energy source to operate BSs. The situation gets worse in low load conditions where the BS remains powered up to provide coverage for few active mobile terminals. The BS switch-off strategies cannot be adopted due to possibility of coverage holes in sparse deployments of BSs in remote areas of the country. In such cases, using renewable energy sources can be more advantageous. In \citep{6731020}, a reference model for renewable energy BS (REBS) has been suggested along with the concept of renewable energy-aware BS. The REBS comprises BS, energy control unit (ECU), and energy sources (renewable and non-renewable). The ECU is the important element that utilizes the energy storage unit in case of excess demand/supply and hence compensates the potential un-reliability of renewable energy sources. However, the presented reference model is very simple and the overall approach does not cover the complexities involved in designing ECU.

The renewable energy sources (solar, wind, fuel cell) are suggested in \citep{6290252} for eco-friendly green 5G cellular networks. In the year 2004, Japanese cell phone operator NTT DOCOMO operated an experimental 3G BS (DoCoMo Eco Tower). This self-powered tower used solar and wind power simultaneously \citep{NTT2004}. In the year 2010, world wide fund for nature (WWF) annual report \citep{WWF2010} was published showing substantial reduction in CO$_2$ emission in China because of using alternative energy sources\footnote{The use of solar and wind energy saved China 48.5 million metric tons of CO$_2$ emissions in the year 2008 and 58.2 million metric tons in the year 2009. Based on the result for China Mobile, and with conservative estimates, 70 million tons of carbon emission reductions had been estimated in the year 2008 which is equivalent to the total CO$_2$ emissions from countries like Sweden, Finland and Norway.}.
In \citep{6102353, 6364971, 6210335}, an energy efficient communication and the dynamics of the smart grid are considered in designing green wireless cellular networks. The author in \citep{6102353} proposed a novel game-theoretical decision making strategy to analyze the impact of smart grid on cellular network. The retailer and consumer are formulated as two players of a Stackelberg game. The proposed decision making scheme considers real time pricing in demand side management mechanism and gives insights into system parameters that affect the retailer's procurement and price decisions. The idea has been extended in \citep{6364971} by considering CoMP to ensure QoS when certain BSs are switched off. Both of these strategies are further extended in \citep{6210335} where service blocking probability is included in the system model. The analysis of the two player game has been enhanced by proving existence as well as uniqueness of the Stackelberg equilibrium. Though the approaches of energy efficient smart grid communication ensures reduction in OPEX and CO$_2$ emissions, the inherent problem of coverage holes due to BS switch-off, more control signaling at air interface due to coupled planes, and much higher backhaul requirements in case of CoMP operation renders such approaches impractical. Moreover, using CoMP to provide coverage for all users of switched-off BS can cause severe blockage and poor QoS. This is because CoMP has originally been designed to ensure cell-edge coverage not for the coverage of all users due to severe backhaul capacity limitations.

The existing approaches for CARC ensures power savings, however, all these approaches have certain shortcomings discussed previously. For example, BS switch-off mechanism in CARC causes coverage holes and in order to provide coverage by the cell range expansion techniques, the transmit power of covering BS increases resulting in inter-cell interference. The existing energy efficient approaches for CARC and the corresponding shortcomings are summarized in Table \ref{Table:Appr_CARC}. The problem of coverage holes and subsequent problem of increased transmit power does not exist in SARC due to inherent ubiquitous coverage of cBS. Similarly, the problem of continuous operation of sparsely deployed BSs in remote locales of the country can best be tackled by providing data services by cBS during off-peak hours. Therefore, SARC can scale with two extreme load conditions (i.e., remote locales and ultra-dense environments).
\begin{table*}[!htb]
\renewcommand{\arraystretch}{1.4}
\caption{Summary of approaches for energy efficient communication.}\label{Table:Appr_CARC}
\vspace{2mm}
\centering
\begin{tabular}{||l||p{4.5 cm}||p{7 cm}||l||}
\hline
\textbf{General Approach} 						& \textbf{Proposed Technique} 													& \textbf{Shortcomings} 																					& \textbf{Ref.}              			\\ \hline \hline
\multirow{10}{*}{Dynamic BS Switch-off} 				& Reduction in activity probability and cooperation/coordination 								& Not suitable for already deployed BSs \newline High backhaul capacity requirements                       									& \citep{6629715}    				\\ \cline{2-4} 
                                        							& Probabilistic sleeping mechanism												& Mobility impact not considered																				& \citep{7037473}				\\ \cline{2-4} 
                                        							& Cell range expansion														& Inter-cell interference due to high transmit power																	& \citep{5208045,5300273}		\\ \cline{2-4} 
                                        							& First/second order statistics of traffic profile											& Not suitable for varying traffic patterns																		& \citep{5683654}				\\ \cline{2-4} 
                                        							& Small cell deployment														& Scales poorly with number of small cell deployments																& \citep{5360741,rost201011}		\\ \cline{2-4} 
                                        							& Temporal and spatial variations of traffic profile										& Suitable for homogeneous deployment																		& \citep{6489498}				\\ \cline{2-4} 
                                        							& Flow level dynamics														& Not suitable for fast fading channels\newline Assumption of Gaussian-noise like inter-cell interference									& \citep{5992823}				\\ \cline{1-4} 
Renewable Energy Resources             					& Alternate energy as main source                             										& \multirow{2}{*}{Not addressing problem of under-utilized network resources}                                                                						& \citep{6731020,6290252,NTT2004}	\\ \cline{1-2} \cline{4-4} 
Smart Grid                         							& Game theoretical approach													&                                                                                                                                             											& \citep{6102353, 6364971, 6210335} 	\\ \hline
\end{tabular}
\vspace{-2mm}
\end{table*}
\subsection{System Capacity}\label{sec:sys_cap}
In the past, voice services dominated data services due to which the cellular systems were mainly designed for the voice traffic. Such systems offered very low system capacity complaint to the capacity requirements of voice services at that time. In the year 2009, the mobile data overtook the voice traffic in terms of total traffic generated on the network. With the emergence of mobile data services, the capacity requirements increased and the total worldwide mobile traffic is now expected to reach very high numbers. A brief view on number of worldwide mobile subscribers excluding WiFi traffic off-loading and including M2M communication \citep{UMTS_2011} is shown in Table \ref{Table:tab1}.
\begin{table}[b]
\vspace{-6mm}
\renewcommand{\arraystretch}{1.25}
    \centering
    \caption{Traffic Forcast}\label{Table:tab1}
\vspace{2mm}
    \begin{tabular}{|c|c|c|c|l|}
    \hline
    \textbf{Category/Year} & \textbf{2010} & \textbf{2015} & \textbf{2020} \\
    \hline
    Global Mobile Subscribers (Million) & 5328 & 7490 & 9684 \\
    \hline
    Total Mobile Traffic (EB) & ~4 & ~45 & ~127 \\
    \hline
    \end{tabular}
\end{table}

The capacity requirements in terms of average area throughput for future mobile networks beyond international mobile telecommunications-advanced (IMT-Advanced)\nomenclature{IMT-Advanced}{International Mobile Telecommunications-Advanced} are studied to be 25 Gb/s/Km$^2$ \citep{5706319} with peak data rate of 4.5 Gb/s/cell in downlink and 2.5 Gb/s/cell in uplink. The spectrum and bandwidth requirements for future IMT-2000 and IMT-Advanced are presented in \citep{IMT}. Such high requirements and explosive growth of mobile data require huge system capacity. In literature, mainly three approaches are considered to meet the capacity requirements. These include spectrum efficiency, spectrum aggregation \citep{6881734}, and network densification \citep{6477646}. The same has been identified by DOCOMO as "The Cube" for future 5G systems \citep{DOCOMO}. The spectrum efficiency targets the capacity enhancements by considering CB, multi-user MIMO, and CoMP. The spectrum aggregation includes carrier aggregation either contiguous or non-contiguous to meet the capacity requirements of different applications. However, the spectrum efficiency/aggregation have a local scope as compared to network densification that has been globally accepted as the cost-effective and agile solution to meet the capacity demands of future cellular systems. The huge number of small cell deployment results in heterogeneity in the network. This heterogeneity is expected to increase in the future by the increased number of D2D and M2M communications. In such ultra-dense HetNet, virtually a personal cell might be required in future to meet the capacity and coverage requirements. The idea of personal cell has been introduced as pCell technology by Artemis Networks \citep{Artemis} where each wireless device will be provided the full bandwidth even in high load conditions and hence each mobile device will have virtually a dedicated personal BS. However, the pCell technology has yet not been commercialized.

In SARC, the capacity enhancements can be realized flexibly. For example, the spectral efficiency can be higher due to reduced control signaling interference. In CARC, every BS is responsible to provide control signaling as well as data services in its coverage area. Therefore, there will be as many control signaling interferers as there are BSs in specific area. In contrast to CARC, smaller number of cBSs will provide global coverage and hence control signaling interferers will be reduced in SARC. Moreover, due to sleeping dBSs, there will be reduced inter-cell interference in data plane. The beam forming and CoMP can be realized centrally at cBS. The adaptive dBS clustering for CoMP operation can be flexible by considering cell-sleeping into account (which is not possible in CARC). By having global coverage of the cBS, traffic off-loading may be realized by establishing D2D communication for common content exchange. 

The network densification in SARC includes deployment of dBSs in the coverage of cBS. In order to enhance capacity, dBSs can be deployed at higher frequency bands with much more bandwidth. In this context, huge bandwidth at mm-Wave spectrum is an attractive choice for high-rate data transmissions \citep{6732923}. In \citep{4457895}, the authors provide detailed design trade-offs and performance requirements to support wireless communication at 60 GHz frequency. The challenges associated with data transmission at this frequency include poor propagation, blocking/shadowing, atmospheric and rain effects \citep{1491267,6824752}. In order to model mm-Wave channel and analyze access performance, ray optics techniques have been used \citep{6134693}. The ray tracing simulations at 72 GHz show that the propagation at such a high frequency can be approximated with limited diffraction and scattering phenomenon. The agreement between channel model and the measurement at mm-Wave band can also be observed in \citep{6253227}. In \citep{6824972}, an air interface design, based on null cyclic prefix single carrier, has been proposed. The ray tracing results and the propagation measurements at mm-Wave show that it is the best candidate for communication at this frequency. The measurement results at mm-Wave (28 and 38 GHz) spectrum with steerable directional antennas are presented in \citep{6515173}. The novel hybrid beamforming scheme and mm-Wave prototyping for indoor and outdoor environment \citep{6736750} asserts the feasibility of wireless communication at this frequency band. All these studies ensure that mm-Wave spectrum has potential gain to ensure high data rate transmission for dBSs in future 5G cellular networks.

In future cellular communication, the mobile devices require several changes from the view point of hardware, software/firmware design, and protocol stack. Though existing smartphone and mobile terminals are multiple random access technology (multi-RAT) capable, however, large antenna array in small form factor is indispensable for mm-Wave transmission. In order to operate in multiple scenarios (e.g., high/low mobility, under legacy network coverage, as relay node, or operating as D2D underlay node, etc), dynamic radio frame and corresponding protocol stack is required for 5G mobile devices. The reader may refer \citep{4784727,tudzarov_protocols_2011} for further details.

The smartphones and mobile devices for 5G networks are introduced as NanoEquipment (NE) in \citep{Nanocore}. The author has discussed 5G RAN and 5G mobile device (i.e., NE) from the perspective of nanocore technology. Using this technology, large antenna array in small form factor can be realized to meet the requirements of the data plane for mm-Wave communication. For high mobility users, low-rate data services may be provided at lower frequencies. Since, control and data planes are expected to operate at different frequencies (i.e., lower frequencies for control/low-rate data and mm-Wave for data plane), dedicated RF chains for control and data planes are mandatory. These are few hardware changes which we can expect for 5G mobile devices. The software and protocol changes are expected to be transparent to mobile devices due to futuristic SDN approach. To realize future cellular communication, dual connectivity for control/data planes and multi-RAT technology for seamless communication (at any available legacy or new air interface) will require self-organized sophisticated radio frame and protocol stack.

In SARC, network densification of dBSs at mm-Wave spectrum can be achieved to meet the capacity requirements. Since, the capacity requirements of future ultra-dense environment are much higher, therefore assigning physical cell identification (PID)\footnote{In LTE/LTE-A, physical cell ID (PCI) is used by UEs to differentiate between neighboring cells and perform signal strength measurements.}\nomenclature{PID}{Physical Cell Identification} to each active/sleeping dBS can be quite challenging. The PID is the physical ID of the cell which is required by UEs to uniquely identify the serving dBS and acquire time/slot synchronization. In current LTE systems, the cell ID can be calculated during initial cell search using primary and secondary synchronization signals (PSS/SSS)\nomenclature{PSS}{Primary Synchronization Signal}\nomenclature{SSS}{Secondary Synchronization Signal}. If the under-utilized dBS is set to sleep mode by the cBS, UE can never know the presence of near-by sleeping dBS. The dBS localization, waking-up and assignment of cell ID, introduced as PID management, are the responsibilities of cBS. Since, cell sleeping is a rare phenomenon in CARC, therefore static PID assignment to always running BSs is a feasible strategy. In SARC, the simple conventional solution of static PID assignment will result into inefficient PID utilization. Many cell IDs will be unused in case of large number of sleeping dBSs. The optimum PID management in SARC can follow on-demand PID assignment in a self-organized manner. This strategy can scale well in case more dBSs are deployed to meet capacity demands. However, this solution comes at the expense of centralized PID management and tight synchronization. For active UEs, the cBS will not only localize the near-by dBS but also assign the PID (in case of sleeping dBS); hence assisting the required time synchronization between dBS and active UEs. Once, the sleeping dBS is active, it can use the assigned ID and corresponding PSS/SSS to provide time/slot synchronization to UEs. In spite of complex processing, the centralized PID management in SARC can bring self-organization which is indispensable for sleeping dBSs in future cellular networks.

Another perspective to meet capacity requirement is to select optimal dBS for data services. Since, cBS has global context information (e.g., positions of dBSs and UEs), it can use simple path-loss, statistical CSI, and load conditions to associate UE to the optimal dBS. Using this simple strategy, the UE can be handed over to the dBS with highest capacity provision. This can be possible because cBS has global knowledge of the coverage area, however, the optimal dBS selection can be challenging due to possibility of cell-sleeping. In such a scenario, cBS has to initiate wake-up mechanisms, assign PID, arrange initial synchronization, and  handover UE to the dBS. Once a successful handover is accomplished, the reduced flow control (minimum required control signaling) with the dBS as compared to the full flow control in CARC, offers a higher degree of freedom to achieve higher data rate. The inherent benefit of SARC architecture is the reduced control signaling in radio frame of dBS. For example, in current LTE and LTE-A systems, the radio frame is 10ms where control signaling is required to be sent periodically along with requested data. Thus, control signaling takes substantial portion of radio frame to provide connection establishment, handover mechanism, and other control procedures. Such restriction does not hold for dBS as majority of the control signaling will be provided by cBS via dual connectivity mechanism. Since control signaling is reduced to minimum in dBS of SARC, the frame size can carry maximum data traffic to meet higher capacity demands.
\subsection{Interference Management}\label{sec:imanag}
In future cellular systems, interference management will be a real challenge due to heterogeneity (small cells, remote radio head, D2D, M2M, multi-RAT services etc.), dense spectrum reuse (overlay/underlay D2D, M2M), and network densification. Although this hierarchical heterogeneity promises tremendous capacity and coverage enhancements, the resulting interference will be manifolds higher as compared to present deployments.

In present OFDMA based cellular systems, intra-cell interference is mitigated using orthogonal sub-carriers, however, inter-cell interference exists due to frequency reuse (reuse-1). This inter-cell interference has negligible effect on cell-centre users and severe effects on cell-edge users. In literature, this inter-cell interference is addressed using different mitigation techniques like randomization, cancellation, and coordination \citep{5441362}. The randomization techniques average out the interference across the whole spectrum using scrambling, interleaving etc. Hence, the interference is not mitigated rather distributed equally and fairly over the system bandwidth \citep{6392819}. The cancellation techniques apply advanced signal processing at the receiver (e.g., interference rejection combining (IRC)) to reject the interference in a single-cell environment whereas coordination techniques push the interference to the cluster level comprising multicell environment. Hence, the notion of interference for cooperative networks has been changed to inter-cluster instead of inter-cell interference. In ideal coordination techniques, intra-cluster interference is completely removed, whereas inter-cluster interference limits the system performance. The ICIC techniques employ either selective frequency reuse, selective power reuse or selective invert power frequency reuse. The frequency reuse (fractional frequency reuse (FFR), partial frequency reuse (PFR), and soft frequency reuse (SFR)) improves the cell-edge performance, however, the major drawback of such techniques is the spectrum under-utilization that directly degrades the overall system performance. The selective power reuse technique is based on higher power for cell-edge users as compared to cell-centre users by keeping orthogonal frequencies to avoid inter-cell collision. This approach overcomes the spectrum under-utilization, however, no significant capacity gain is achievable since interference avoidance in this case is entirely dependent on good channel conditions \citep{4907410}. The invert power frequency reuse technique is the hybrid of frequency and power reuse. This technique is suggested to achieve performance trade-off between under-utilized spectrum with higher cell-edge throughput and fully-utilized spectrum with lower cell-edge throughput \citep{5506110,5450256}.

The ICIC techniques have further been evolved as eICIC for LTE-A. These techniques are categorized into time, frequency, and power domains. In time domain, the interference is handled by sending either ABS or employing symbol shift for the two interfering cells (aggressor and victim) \citep{6475212}. The cell selection bias is introduced to ensure received signal strength based user association in favor of picocell. In frequency domain, the physical signals and control channels are completely orthogonal among aggressor and victim cells, thereby mitigating interference at the cost of reduced bandwidth \citep{D_Lopez2011}. In order to optimize the resources and employ interference control, almost blank resource block (ABRB) \nomenclature{ABRB}{Almost Blank Resource Block} is suggested in \citep{6725662}. The ABRB is defined over both time and frequency domains unlike simple time-domain ABS approach. Hence, it provides more granularity in resource allocation. The ABRB is a generalization of ABS approach and it provides further improvements by providing co-tier (macro-macro) interference control along with cross-tier (macro-pico) interference control. The power domain techniques employ power control mechanism in indoor low-power nodes, however, reducing the maximum transmit power of low-power nodes may degrade the overall performance especially in case of femtocells \citep{6392819}.

In SARC, the interference management has some potential flexibility. Since, the cBS and dBSs are operated on different bands, hence the cBS UEs can roam even in the coverage area of dBSs without causing interference. In the coverage area of cBS, the cell-centre cBS UEs will see no interference from neighboring cBSs due to longer path-loss. However, the cell-edge cBS UEs will be effected by the inter-cell interference. In case of dBS UEs, inter-cell interference will be higher due to ultra-dense deployments of dBSs. In such dense environment, interference-aware transmission may be realized at two levels. The first may consider interference mitigation between cBSs by realizing long-range cBS clusters. The second may consider CB for clusters of densely deployed dBSs.
\subsection{Mobility Management}\label{sec:mob_man}
The optimal mobility management ensures the capacity and coverage of mobile cellular networks. In literature, different approaches are reported for mobility management. In \citep{7022987}, macrocell cooperation and Manhattan grid layout has been proposed for mobility management. The simulation results show that without this cooperation, dense small cells would require at least 4 times more re-connection load. This study investigates potential advantages in mobility management due to macrocell cooperation, however, cell sleeping phenomenon has not been considered in this study. The authors in \citep{7059729} analyzed the mean handover rate and the mean sojourn time in macrocell assisted small cell architecture. The BSs are deployed as poisson point process (PPP) with serving zones as poisson voronoi tessellations (PVT). The random waypoint is considered for the user mobility. The analytic expression show that the handover rate and sojourn time are simply a function of user velocity, transmission probability, and BS density, however, similar to \citep{7022987}, cell sleeping phenomenon has not been considered. The study in \citep{6477646} considered 4 Phantom cells per macrocell to evaluate the handover performance. It has been observed that the handover failure gets worse as the density of phantom cells increases. The handovers are only considered for Phantom-Cell-to-Phantom-Cell without investigating the impact of macrocell handovers and cell sleeping on overall mobility management.

The potential advantages of SDN technology for mobility management has been discussed in \citep{7063430}. The functional description of three approaches for handover management are presented. These include 1) Centralized SDN, 2) Semi-centralized SDN, and 3) Hierarchical SDN. The main problem of preserving session continuity and scalability of handovers is discussed. Similarly, the functional description and architecture of DMM in SDN/OpenFlow has been presented in \citep{7063379}. Both these studies are mainly a functional level discussion without providing design guidelines for mobility management. For further details on SDN networking, the reader is referred to a recent survey \citep{6994333}.

The mobility management for high-speed railway wireless communication networks has been considered in \citep{6807723, 6710300}. The evolution of GSM for railway (GSM-R) to LTE for railway (LTE-R) has yet not been standardized \citep{choi_standards_2014}, however, LTE has been considered to study the impact on performance of European Train Control System (ETCS) railway signaling \citep{6807723}. This study did not consider control and data planes separation unlike \citep{6710300} where theoretical analysis and simulation results are presented to emphasize higher security of train control system and larger capacity for passenger services by using separation framework. This study, however, did not mention the impact of fast handovers on system performance.

\begin{figure}[!htb]
\centering
  \includegraphics[width = 1\columnwidth]{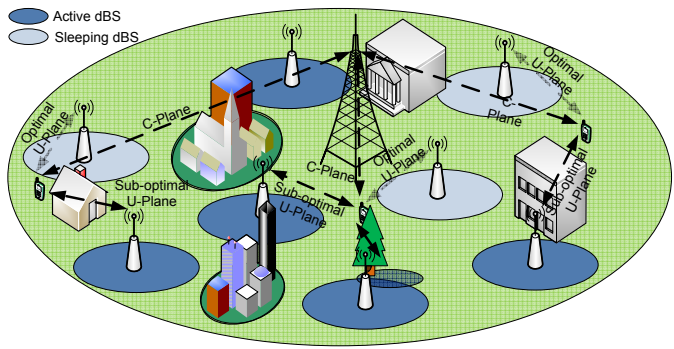}
  \caption{Active and sleeping dBSs in coverage area of cBS.}\label{Figure:ss}
\end{figure}

In future cellular systems, mobility management should be as seamless as possible for multicell and multi-RAT technologies in order to provide ubiquitous coverage and meet the capacity demands of UEs. In CARC, coverage is provided distributively by different BSs in their respective coverage areas. In such architecture, each BS has limited local knowledge of the network. Since, the coverage and control is not centralized, hence the serving BS might not be knowing the possible optimal sleeping BS in the vicinity to initiate handover. It might be the case that the serving BS initiate handover to the first tier of neighboring sub-optimal multicell or multi-RAT BS whereas the optimal sleeping BS is present in the higher tiers of neighboring BSs. This is due to the non-availability of signal strength of sleeping BSs as well as lack of global knowledge of the multiple tiers of BSs. In SARC, the cBS is centralized with global knowledge of all dBSs (active or sleeping) in multiple tiers in the whole coverage area. The cBS will have the context information and it can predict the signal strength of sleeping dBS e.g., using simple distance dependent path-loss and statistical channel conditions, to select, awake and initiate handover which is not possible in CARC. However, this flexibility comes at the expense of more intense signal processing in cBS to find the optimal set of BSs for handover procedures. One of the limitation of SARC is that if cBS fails due to any reason, the whole coverage area might black out, whereas in CARC, failure of one BS just affects a small portion of the coverage area. However, this limitation can be mitigated by a self-organized backup cBS. The global coverage with sleeping data cells in coverage area and optimal/sub-optimal dBS selection scenarios are depicted in Fig. \ref{Figure:ss}.

In this figure, the channel conditions vary for different deployment scenarios. It might be possible that the closest dBS provides worst data service due to deep fades and shadowing effects as compared to the farthest sleeping dBS. However, the central control of cBS provides flexibility in optimal/sub-optimal dBS selection by incorporating wake up mechanism for sleeping dBSs.
\subsubsection{Cell (re) Selection}
The cell (re)selection in SARC is different from CARC. Two possible scenarios of UE activity are described. The first comprises in-active mode where UE does not require data services. Only coverage is required which is provided by the cBS. The UE can be stationary or moving in the coverage area of either dBSs or cBS without requiring any handover procedures. In CARC, the cell (re)selection and subsequent handover is carried out for even in-active UEs to provide ubiquitous coverage. This is due to the coupling of control and data planes. The second case consists of active mode where UE requires data services. This case is quite complex as compared to CARC where data session needs to be established in the same BS that is providing the control signaling. In CARC, if the BS is sleeping, then the coverage has to be provided by neighboring BSs whereas in SARC, cell-sleeping is more flexible as global coverage is provided by the cBS (even in coverage areas of dBSs). Therefore, the notion of cell sleeping in SARC is different than CARC. Since the SARC is more feasible for cell sleeping mechanisms, the cell (re)selection procedures become complex as compared to CARC. The mobility management in SARC has to consider the sleeping cells into account while optimizing the dBS selection for the requested data service. Although cBS has global knowledge of dBSs and UEs in the coverage area, it does not know the channel conditions between sleeping dBS and associated near-by UEs. In best channel conditions, the near-by sleeping cell is the best candidate for data services so the cBS can wake up the sleeping cell by assuming simple path-loss model and performs the cell (re)selection as well as handover procedures. However, in worst channel conditions, there can be a case that the near-by dBS might provide worse received signal strength as compared to a far sleeping dBS. In CARC, the UE reports RSRP measurements of neighboring BSs to the serving BS. In SARC, the cBS is long range and hence UE can report the RSRP measurements of dBSs directly to the cBS that can manage cell (re)selection globally (further potential gains of this strategy are highlighted in Sec. \ref{sec:cb}). However, the measurements by the UE will exclude sleeping dBSs and therefore some other mechanism should be devised for predicting the channel conditions of sleeping dBSs. The cell (re)selection may be based on conventional procedures (RSRP based), though, such procedures are more challenging in SARC due to the sleeping cells in coverage area. The complexity is traded-off with more centralized control on cell-sleeping to conserve energy for green cellular communication.
\subsubsection{Handover Procedures}\label{MM:ho_proc}
\begin{figure}[h]
\centering
  \includegraphics[width = 1\columnwidth,height = 2.75in]{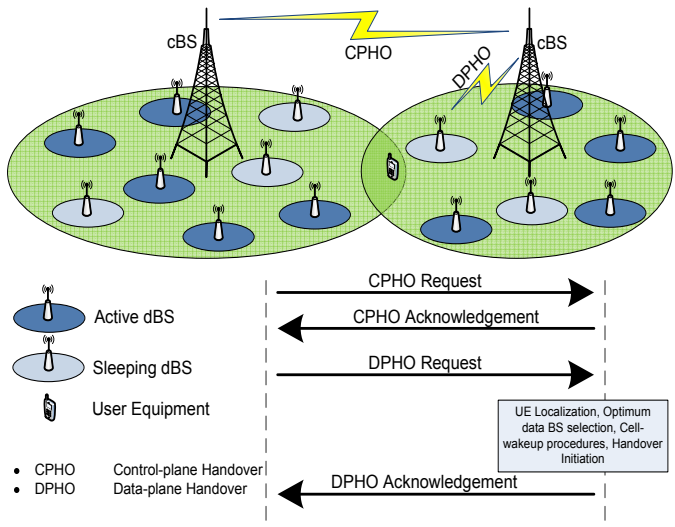}
  \caption{Handover Procedure}\label{Figure:ho}
\vspace{-2mm}
\end{figure}

The handover requirements in SARC and CARC are different. In CARC, complete handover is initiated for cell-centre or cell-edge users. However, in SARC, partial handovers might be required depending on cell-centre or cell-edge users. Therefore, SARC handovers can be classified into partial and full handovers. In partial handover, only data plane handover (DPHO) is required for cell-centre users and control plane is intact. For cell-edge users, complete handover consisting of control plane handover (CPHO) and DPHO might be required. Therefore, SARC offers significant reduction in CPHO overheads for cell-centre users. However, the complete handover in SARC is complex as compared to CARC where the handover is performed softly as both control and data sessions are handed over to a single neighboring BS. In SARC, the control handover is made to the neighboring cBS and data sessions are handed over to the active or sleeping dBSs in the neighboring coverage area. This procedure might produce delays in case of sleeping dBSs. Therefore, an agile and robust soft handover is needed in SARC for cell-edge users. In spite of this complexity, the potential gains in SARC due to CPHO overhead reduction can be significant due to the increased number of cell-centre users. The simplest handover procedure for CPHO and DPHO is depicted in Fig. \ref{Figure:ho}.
In this figure, we can have different scenarios for handovers which are discussed below.
\paragraph{Control Plane Handover (CPHO)}
In CPHO, the cell-centre users do not require handover as discussed above. However, for cell-edge users, CPHO is required whether the UE is in active or in-active data sessions. In case of in-active data sessions, the CPHO is not complex and it can be initiated and performed quickly without incurring delays and hence providing ubiquitous coverage. However, in active data sessions, handover procedures might be challenging due to decoupling of control and data BSs.
\begin{table}[!htb]
\renewcommand{\arraystretch}{1.5}
\centering
\caption{Mobility Management in CARC and SARC.}\label{Table:mm}
\begin{tcolorbox}[tab4, title = Quick View of Mobility Management, boxrule=0.3mm,top=0.3mm,bottom=0.3mm,left=0.3mm,right=0.3mm,
rightrule=0.3mm]

\begin{tcolorbox}[tab5,tabularx={>{\raggedright\arraybackslash}p{0.8in}||X|X}, boxrule=0.25mm,top=0.25mm,bottom=0.25mm,left=0.25mm,right=0.25mm,
rightrule=0.25mm]
			& \bf CARC									& \bf SARC	\\ \hline
\bf BS Knowledge	& Knowledge of the multiple tiers of BSs is not available.		& Global knowledge of all dBSs (active or sleeping) due to centralized cBS	\\ \hline 
\bf Sleeping BSs	& RSRP is not available.							& cBS can predict statistical RSRP.								\\ \hline
\bf Handover	& Serving BS cannot initiate handover to sleeping BS.				& Initiate wake-up mechanism and subsequent handover			\\ \hline
\bf BS failure 	& Failure of one BS affects small portion of the coverage area.	& Whole coverage area can be affected.
\end{tcolorbox}
\begin{tcolorbox}[tab5, tabularx = {X||X}, title=Cell (re)selection, boxrule=0.25mm,top=0.25mm,bottom=0.25mm,left=0.25mm,right=0.25mm,
rightrule=0.25mm]
\bf CARC									& \bf SARC	\\ \hline
If the BS is sleeping, then the coverage has to be provided by neighboring BSs.			& Flexible cell sleeping mechanisms require complex cell (re)selection procedures.			\\ \hline
UE reports RSRP measurements of neighboring BSs to the serving BS. 	& UE reports the RSRP measurements of dBSs directly to the cBS that can manage cell (re)selection globally.
\end{tcolorbox}
\begin{tcolorbox}[tab5, tabularx = {X||X}, title= Handover Procedures, boxrule=0.25mm,top=0.25mm,bottom=0.25mm,left=0.25mm,right=0.25mm,
rightrule=0.25mm]
Complete handover is initiated for cell-center or cell-edge users 	& Partial handover for cell-center users and full handovers for cell-edge users 							\\ \hline
Complete handover is performed softly as control/data sessions are handed over to a single neighboring BS. & Complete handover in SARC is complex due to decoupled control and data planes.
\end{tcolorbox}
\end{tcolorbox}
\vspace{-4mm}
\end{table}

\begin{table*}[!htb]
\makeatletter
\newcommand*{\compress}{\@minipagetrue}
\makeatother
\renewcommand{\arraystretch}{1.25}
\caption{Shortcomings/potential gains due to CARC/SARC.}\label{Table:Short_Gains}
\vspace{2mm}
\centering
\begin{tcolorbox}[tab3,tabularx={>{\raggedright\arraybackslash}p{0.6in}||>{\raggedright\arraybackslash}p{1.3in}||X||X}]
\textbf{Perf. Measure} 							& \textbf{General Approach} 						& \textbf{Shortcomings due to CARC} 															& \textbf{Potential Gains due to SARC}              			\\ \hline\hline
\multirow{8}{*}{Energy} \multirow{8}{*}{Efficiency}			
											& \compress \begin{itemize}[leftmargin=1.25em]
												\renewcommand{\labelitemi}{$\Rightarrow$}
												\vspace*{6mm} 
												\item Dynamic BS switch-off 
												\item Renewable energy resources 
												\item Smart grid
												\vspace*{-\baselineskip} 
											\end{itemize} 																					
																					& \compress \begin{itemize}[leftmargin=1.25em]
																						\renewcommand{\labelitemi}{$\Rightarrow$}
																						\item Generation of coverage holes due to BS switch-off.
																						\item Higher interference due to increased transmit power in cell range expansion.
																						\item Higher overhead/HO signaling results in more power consumption. 
																						\item Renewable and smart grid resources do not address the problem of under-utilized network resources rather provide alternate source of energy.
																						\vspace*{-\baselineskip}
																					\end{itemize}																		&\compress\begin{itemize}[leftmargin=1.25em]
																																										\renewcommand{\labelitemi}{$\Rightarrow$}
																																										\item No coverage holes due to ubiquitous coverage.
																																										\item No cell range expansion is required due to ubiquitous coverage.
																																										\item Reduced power consumption due to reduced overhead/HO signaling.
																																										\item Renewable and smart grid resources can be added in main power source.
																																										\vspace*{-\baselineskip}
																																									\end{itemize} 						\\ \hline
\multirow{11}{*}{System} \multirow{11}{*}{Capacity}
											& \compress \begin{itemize}[leftmargin=1.25em]
												\renewcommand{\labelitemi}{$\Rightarrow$}
												\vspace*{15mm} 
												\item Spectrum efficiency
												\item Spectrum aggregation 
												\item Network densification 
												\vspace*{-\baselineskip} 
											\end{itemize} 																		
																					&\compress\begin{itemize}[leftmargin=1.25em]
																						\renewcommand{\labelitemi}{$\Rightarrow$}
																						\item High frequency (mm-Wave) cannot be used for data services since it will result in reduced coverage (due to spot beams).
																						\item Since mm-Wave communication cannot be realized, the huge contiguous spectrum cannot be utilized for capacity enhancements.
																						\item Control/overhead signaling restricts payload size resulting in reduced system capacity.
																						\item Limited pro-active caching for cooperation set due to distributed and local context.
																						\vspace*{-\baselineskip} 
																					\end{itemize}														
																																									& \compress\begin{itemize}[leftmargin=1.25em]
																																										\renewcommand{\labelitemi}{$\Rightarrow$}
																																										\item High frequency (mm-Wave) communication to enhance system capacity without loosing coverage.
																																										\item The huge contiguous spectrum can be utilized for very high rate data services.
																																										\item Control is decoupled, therefore, radio frame may contain maximum payload size.
																																										\item Global context allows wake-up mechanism as well as pro-active caching for cooperation.
																																										\vspace*{-\baselineskip}
																																									\end{itemize}						\\ \hline 
\multirow{6}{*}{Interference} \multirow{6}{*}{Management}
											& \compress \begin{itemize}[leftmargin=1.25em]
												\renewcommand{\labelitemi}{$\Rightarrow$}
												\vspace*{6mm} 
												\item Interference avoidance
												\item ICIC/eICIC 
												\item CB and CoMP 
												\vspace*{-\baselineskip} 
											\end{itemize} 
																					&\compress\begin{itemize}[leftmargin=1.25em]
																						\renewcommand{\labelitemi}{$\Rightarrow$}
																						\item Low degree of freedom for active/inactive users
			     																			\item While moving, in-active users require handovers that causes interference.
			     																			\item Active users require full handover (CPHO+DPHO)
			     																			\item Ping-pong effect due to high mobility near cell-edge.
																						\vspace*{-\baselineskip}
																					\end{itemize}														
																																									& \compress\begin{itemize}[leftmargin=1.25em]
																																										\renewcommand{\labelitemi}{$\Rightarrow$}
																																										\item No CPHO for in-active users.
     																																										\item Only DPHO for active users.
     																																										\item Reduced interference due to less CPHO and DPHO.
     																																										\item Ping-pong effect can be controlled by providing data services via control plane.
																																										\vspace*{-\baselineskip}
																																									\end{itemize}						\\ \hline 
\multirow{6}{*}{Mobility} \multirow{6}{*}{Management}
											& \compress \begin{itemize}[leftmargin=1.25em]
												\renewcommand{\labelitemi}{$\Rightarrow$} 
												\vspace*{4mm} 
												\item Macrocell cooperation
												\item Macrocell assisted small cells 
												\item SDN approach
												\vspace*{-\baselineskip} 
											\end{itemize}
																					&\compress\begin{itemize}[leftmargin=1.25em]
																						\renewcommand{\labelitemi}{$\Rightarrow$}
																						\item Cell sleeping is not flexible.
			      																			\item CSI acquisition is challenging in sleeping cell scenario.
			      																			\item Cooperation requires overhead signaling on backhaul. 
																						\vspace*{-\baselineskip}
																					\end{itemize}														
																																									& \compress\begin{itemize}[leftmargin=1.25em]
																																										\renewcommand{\labelitemi}{$\Rightarrow$}
																																										\item Mobility management is ensured by always-running cBS with flexible dBSs sleeping possibility.
      																																										\item Centralized control of SARC may exploit statistical CSI for sleeping cells.
      																																										\item Ubiquitous control in SARC may exploit pro-active network caching to overcome overhead signaling on backhaul.
																																										\vspace*{-\baselineskip}
																																									\end{itemize}						\\  
\end{tcolorbox}
\vspace{-0.5mm}
\end{table*}
\paragraph{Data Plane Handover (DPHO)}
In current LTE systems, the handover is initiated based on events (A1-A5) where the main theme behind these events is to set a certain threshold between serving and neighboring BSs for handover initiation. When the signal strength of neighboring cell is higher than the serving cell, a handover procedure is initiated. A brief description of events \citep{3gpp.36.331} is described as follows:-

\begin{itemize}
  \item \textbf{A1}:    Serving cell becomes better than threshold.
  \item \textbf{A2}:    Serving cell becomes worse than threshold.
  \item \textbf{A3}:    Neighbor becomes offset dB better than serving cell.
  \item \textbf{A4}:    Neighbor becomes better than threshold.
  \item \textbf{A5}:    Serving becomes worse than threshold 1 and neighbor becomes better than threshold 2.
\end{itemize}

In SARC, the cell-centre users will require DPHO based on UE activity. This case can be quite complex as compared to CARC. In case the dBSs are active, then the DPHO procedures are not very complex and no waking-up mechanisms are required. However, the sleeping dBSs may pose challenge to the cBS. Since in this case, the cBS may require localization of UE and prediction of sleeping dBSs channel conditions for optimum dBS selection among the neighboring dBSs. Many factors need to be considered before actually waking-up the dBS. It might not be advisable to wake-up a dBS just for short-time roaming users. In order to provide optimum DPHO, mobility trends might be predicted based on context and history to differentiate between short/long term UE camping and avoid ping-pong effects before waking up the dBS. The quick view of mobility management, cell (re)selection, and handover procedure is given in Table \ref{Table:mm}.

The discussion of above mentioned performance measures target two aspects: 1) shortcomings of the existing approaches, and 2) potential gains due to SARC architecture. In order to provide a quick view of this discussion, we provide shortcomings of CARC and potential gains due to SARC in Table \ref{Table:Short_Gains}.
\section{Cooperative Communication (Essential Background)}\label{sec:cc}
The cooperative communication is a broad term encompassing mainly two categories of wireless networks e.g., cellular, and ad-hoc. The objective of cooperation in both cases is same i.e., nodes should act as cooperative agents for other nodes in order to improve, for example, coverage probability, interference management, and capacity of the overall system \citep{ahn2011wireless}. However, the cooperation strategies are different due to the presence and absence of infrastructure in former and later cases, respectively. For example, in ad-hoc networks, wireless nodes spontaneously and dynamically self-organize into an arbitrary and temporary infrastructure \citep{rubinstein2006survey} without relying on central controller (e.g., BS) for the signaling flow and connection management, whereas, in cellular cooperative networks, the wireless nodes are controlled and dependent on the serving BSs. However, cooperative communication with little involvement of BS can be seen in case of D2D communication. This type of communication can be considered as infrastructure based ad-hoc links where peers act as either mobile relays (e.g., content dissemination) or the source nodes (e.g., file transfer, exchange of common contents etc) with little involvement (control signaling) of access and core network. In this context, D2D communication can also be categorized as cooperative communication to assist the network for content dissemination or ad-hoc type direct communication.

In this section, we provide essential background to understand infrastructure based multicell BS coordination (e.g., CoMP), self-organized BS clustering and network-controlled D2D communication.
\subsection{CoMP Classification}\label{comp_classification}
\begin{figure*}[!htb]
\centering
  \includegraphics[center, scale = 0.75]{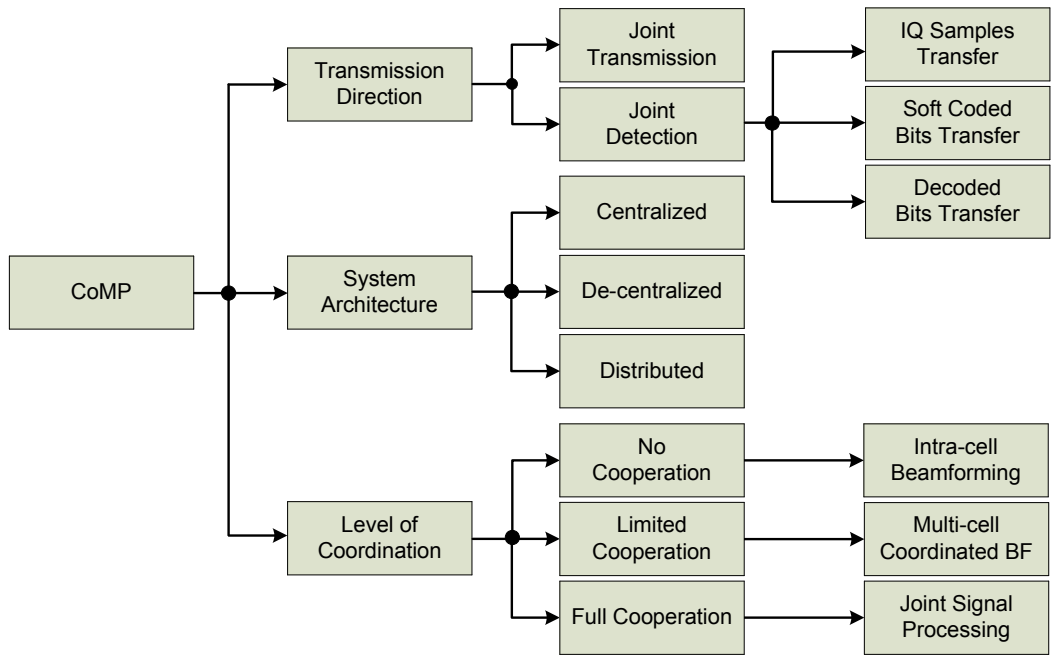}
 \caption{CoMP Classification}\label{Figure:comp_class}
\vspace{-4mm}
\end{figure*}

CoMP can be classified from a number of different perspectives. For example, if the transmission direction is taken into consideration, then CoMP is classified as either joint detection (JD)\nomenclature{JD}{Joint Detection}(uplink) or joint transmission (JT)\nomenclature{JT}{Joint Transmission} (downlink). From the cooperation system architecture, CoMP can either be centralized, decentralized or distributed. The level of CoMP coordination is quantized into no, limited, and full cooperation. Based on this quantization, CoMP scales into either intra-cell beamforming, multicell CB, or fully coordinated CoMP. The classification of CoMP is illustrated in Fig. \ref{Figure:comp_class}.
\subsubsection{Joint Detection}\label{joint_detection}
In multicell JD, each BS receives signals from its respective UEs and exchanges either quantized or un-quantized signal between cooperating BSs. In a typical scenario, a BS suffering from high co-channel interference sends cooperation request to the participating BSs. This request includes the cooperation mode and physical resource block (PRB) associated with the effected UE. The cooperating BSs exchange the quantized signal of the requested UE depending on the cooperation mode as follows:
\begin{itemize}
  \item \textbf{IQ Samples Transfer}:    A frequency-domain in-phase quadrature (IQ) samples representing complex constellation points of  the requested UE are extracted from Fast Fourier Transform (FFT) module and transferred to the serving BS. The serving BS processes the IQ samples as if they were received by its own antennas.
  \item \textbf{Soft Coded Bits Transfer}:    In this cooperation mode, the cooperation request of serving BS must contain not only the PRBs of the transmitted signal, but also its modulation and reference signals. After equalizing and demodulating the received signal, the cooperating BSs transfer the quantized soft values of the coded bits back to the serving BS.
  \item \textbf{Decoded Bits Transfer}:    The serving BS also mentions the decoder of the associated UE and shares it with cooperating BSs which demodulate and decode the signal, perform cyclic redundancy check (CRC)\nomenclature{CRC}{Cyclic Redundancy Check} and transfer the decoded data back to the serving BS on successful CRC check. After receiving the response message, the serving BS performs selection combining. For details on JD algorithms, the reader is referred to \citep{1194444, 738086, 774855, 1271237}.
\end{itemize}
\subsubsection{Joint Transmission}
In multicell JT, CSI and user data of each UE in cooperation set is exchanged between cooperating BSs. Each BS designs beamformers and jointly transmit the data to the target UE. In this scheme, coherent transmission plays a key role to achieve maximum performance gain of JT. The reader is referred to \citep{1207369, 1291726, 4203115} for optimal JT strategies.
\subsubsection{Centralized}
In centralized JD, the cooperating BSs decode the received signal of the corresponding UEs according to the cooperation mode (mentioned in \ref{joint_detection}) and share it with the BS that acts as a centralized node to jointly decode all UEs. In case of centralized JT, each participating BS of cooperation cluster send the CSI to the centralized controller which finds global optimal precoding vectors. These precoding vectors are then shared between participating BSs to exploit inter-cell interference.
\subsubsection{De-centralized}
In decentralized JD, every cooperating BS individually and independently decodes the uplink transmission of respective UEs by exploiting CSI that has been shared between all BSs in the cooperation cluster. In case of JT, every cooperating BS has different extent of CSI knowledge and no BS in the cooperation cluster has full knowledge of global CSI at transmitter. The global optimal precoding is not possible in this case and hence this decentralized JT provides sub-optimal solutions.
\subsubsection{Distributed}
This scheme is similar to the centralized approach with only difference that there is no dedicated central unit (CU)\nomenclature{CU}{Central Unit} and any participating BS can act as a centralized node in a distributed manner.
\subsubsection{Intra-cell Beamforming}
In case of non-cooperative CoMP, BSs do not exchange information, rather perform individual intra-cell beamforming based on limited feedback from their respected UEs. Based on feedback, each serving BS performs interference-aware scheduling and the corresponding UEs have the capability of IRC receiver. 
\subsubsection{multicell Coordinated Beamforming}
In this case, the cooperating BSs exchange CSI between each other in order to reduce the inter-cell interference. This level of coordination requires small backhaul capacity and is known as coordinated beamforming in 3GPP LTE-A literature.
\subsubsection{Full Cooperation}
In case of full cooperation, CSI and user data of each CoMP-enabled UE is exchanged between cooperating BSs. This scheme requires very large backhaul capacity and strict synchronization requirements to perform joint signal processing. The full extent of this cooperation may be exploited by adopting coordinated and coherent transmission to the target UE.
\subsection{CoMP Clustering}\label{sec:comp_clustering}
\begin{figure}[!htb]
\centering
  \includegraphics[width = 1\columnwidth]{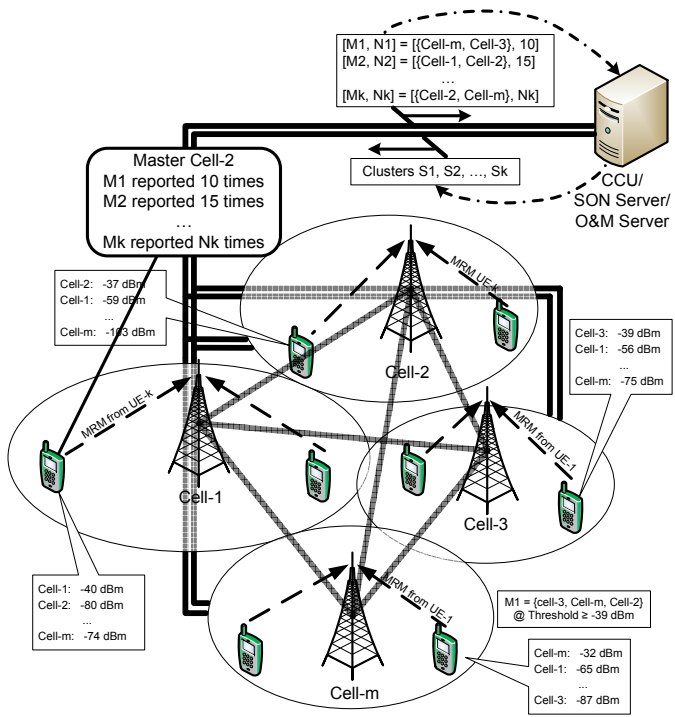}
  \caption{Self-organizing Network Based Adaptive Clustering}\label{Figure:adaptive_clustering}
\end{figure}

Due to signaling overheads on air interface and backhaul, the number of BSs in cooperation cluster is limited in practice. For such cooperating BSs, the clustering can be static or dynamic. The static clustering is designed on the basis of geographical positioning of BSs and is kept constant over time and channel conditions. However, dynamic or adaptive clustering adapts the channel conditions and is comparatively more complex. The adaptive clustering for current cellular system is suggested by exploiting existing RF measurements reported by UEs to the serving BS \citep{garavaglia_adaptive_2014} as shown in Fig. \ref{Figure:adaptive_clustering}. 

In such an approach, huge number of average RSRP measurements are extracted from the measurement report messages by serving BSs of respective UEs. These huge measurements are categorized in the form of reporting sets and sent to the CoMP CU (CCU)\nomenclature{CCU}{CoMP Central Unit} which selects the cooperation cluster based on some performance indicator. These indicators may, for example, include system load, delay, system complexity, combined signal strength, user priority classifications, or other network related metrics \citep{garavaglia_adaptive_2014}. The advantage of this approach is that it can utilize the existing framework of 3GPP (functions such as automatic neighbor relation (ANR), neighbor relation tables (NRTs)) to provide self-organizing network (SON)\nomenclature{SON}{Self-organizing Network} based clustering solution.
\subsection{Decive-to-device Cooperation}\label{sec:d2d_coop}
In present cellular systems, we have HetNets that comprise macrocells, small cells (micro, pico, femto), access points, and smart mobile devices. In future cellular systems, ultra-dense HetNets are expected where capacity and coverage can be met by cooperation between different nodes. In this context, even more smaller granularity of cooperation is expected e.g., CB/CoMP at device level (D2D CoMP) and D2D cooperation for content dissemination or common information exchange.

D2D communication has an old origin in the form of ad-hoc and personal area networking technologies in unlicensed spectrum bands e.g., industrial, scientific, and medical (ISM)\nomenclature{ISM}{Industrial, Scientific, and Medical} bands. In this case, short range communication is possible without infrastructure unlike cellular communication where network control is mandatory. Although such ad-hoc communication requires very less control signaling, it inherits certain drawbacks such as limited content sharing, no point-to-multipoint links, synchronization issues, authentication, and security concerns. D2D communication has also been proposed in licensed spectrum especially in cellular bands in either ad-hoc or network-assisted mode. The ad-hoc mode of D2D communication in licensed spectrum offers limited applications similar to the unlicensed counterpart, however, network-assisted D2D communication in cellular band has many applications and services including proximity-based commercial services, social networking, video sharing, mobile relaying, gaming, traffic offloading, capacity enhancement (frequency reuse), extended cellular coverage, and improved energy efficient communication.

D2D communication has been studied by research community quite long. In early 2006, mobile communication system Aura-Net, based on wireless technology FlashLinQ, was proposed. This communication system exploited D2D communication for proximity-aware inter-networking to enhance and augment the capacity and coverage of wireless wide area network (WWAN) \citep{5675775}. The proposed system features distributed spatial spectrum reuse protocol that is scalable to different levels of proximal granularity. It is mentioned that Aura-Net provides a template for future proximal aware ``Internet of Things''.

The smart communication devices have the capability to be virtually connected to any device, any time, anywhere. This global connectivity offers remoteness as well as proximity at the same time. Coupled with proximity services, the ultra-dense heterogeneity of future cellular networks can be exploited to achieve potential advantages of low-range high-rate D2D data communication to enhance capacity and coverage. D2D communication is considered as a sub-feature of 3GPP  LTE-Direct Rel-12 \citep{balraj_lte_2012}. It comprises two main features:
\begin{enumerate}
\item Device to Device Peer Discovery
\item Device to Device Data Communications
\end{enumerate}

In order to complement huge SC deployments and overcome OPEX and energy efficiency concerns, traffic off-loading from cellular to multi-RAT networks, other unlicensed wireless infrastructures (e.g., WiFi) and multi-hop ad hoc links between devices drew much attention recently. The MOTO project \citep{MOTO} funded by the European Commission under FP7 proposes traffic offloading where D2D communication is one of the ingredients. The establishment of D2D links can be considered as ad-hoc network in infrastructure where the network resources are reused by mobile peers directly with little involvement (control signaling) of access and core network. In this hybrid architecture (infrastructure based ad-hoc links), huge capacity, ubiquitous coverage, energy efficiency, and backhaul gains are promised by exploiting maximum D2D links and reusing the resources optimally.

D2D communication is being considered as an integral part of next generation cellular networks where proximity services and social networks are dominating over conventional services. The network-assisted D2D communication offers another tier of communication within a cell by reusing the spectrum resources. The reduced distance between nodes improves spectral efficiency, throughput per area, energy efficiency, and latency. The link reliability can be improved by migrating from multi-hop to single hop communication (mesh-like topology). The coverage can be enhanced by multi-hop cooperation between devices which can be the only communication in case of no coverage-zone, coverage holes, and emergency situation. The load balancing and load management can be optimized by network and device pro-active caching of common information and offloading the devices to establish direct links \citep{6881261}. Hence, D2D links in future cellular networks are key enablers for traffic off-loading, reducing access delays, optimal resource utilization, capacity and coverage enhancements, and energy efficient communication.

The huge potential performance gains due to direct communication are coupled with certain challenges that include quality-of-experience (QoE), quality-of-protection (QoP), user consent, battery issues, and cellular aspects. These factors are very important and can directly effect the performance gains of D2D communication. The QoE includes user perception, expectations, and experience that needs to be maintained in cellular and direct mode of communication. The QoE is a measure of user's desired or expected experience about cellular services. Though user might not be interested in specific mode of communication (cellular or D2D), he can be considered as perceiving seamless switching between two modes and enjoying services at agreed QoS. The QoP refers to the confidentiality and privacy which is even more severe when the locations and contents may be compromised by intruding D2D partner. However, this can be tackled by incorporating simple authentication, authorization and accounting (AAA) procedures to block such attacks. Even with this solution, using the device without the consent and permission of the mobile owner is a big problem along with battery consumption issues. Using the device for D2D relaying, for example, without incentivizing the mobile owner can not be realized practically. The cellular aspects include interference management due to underlay D2D network, optimal number of D2D nodes, exploitation of common interests (social relationship strength to harness D2D communication), CSI between nodes, and synchronized switching between cellular and D2D nodes.
 
\begin{figure}[!htb]
\centering
\includegraphics[width = 1\columnwidth]{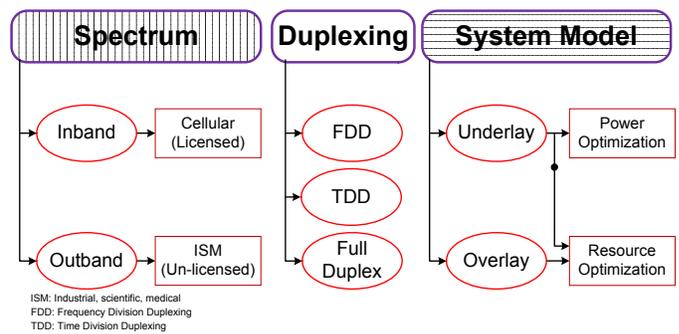}
\caption{D2D Taxonomy}\label{Figure:d2d_class}
\vspace{-5mm}
\end{figure}

D2D communication can be classified in a taxonomic representation as shown in Fig. \ref{Figure:d2d_class}.

\begin{figure*}[!htb]
    \centering
    \subfigure[Conventional Architecture (CARC)]
    {
        \includegraphics[width = 0.96\columnwidth, height = 2.75in]{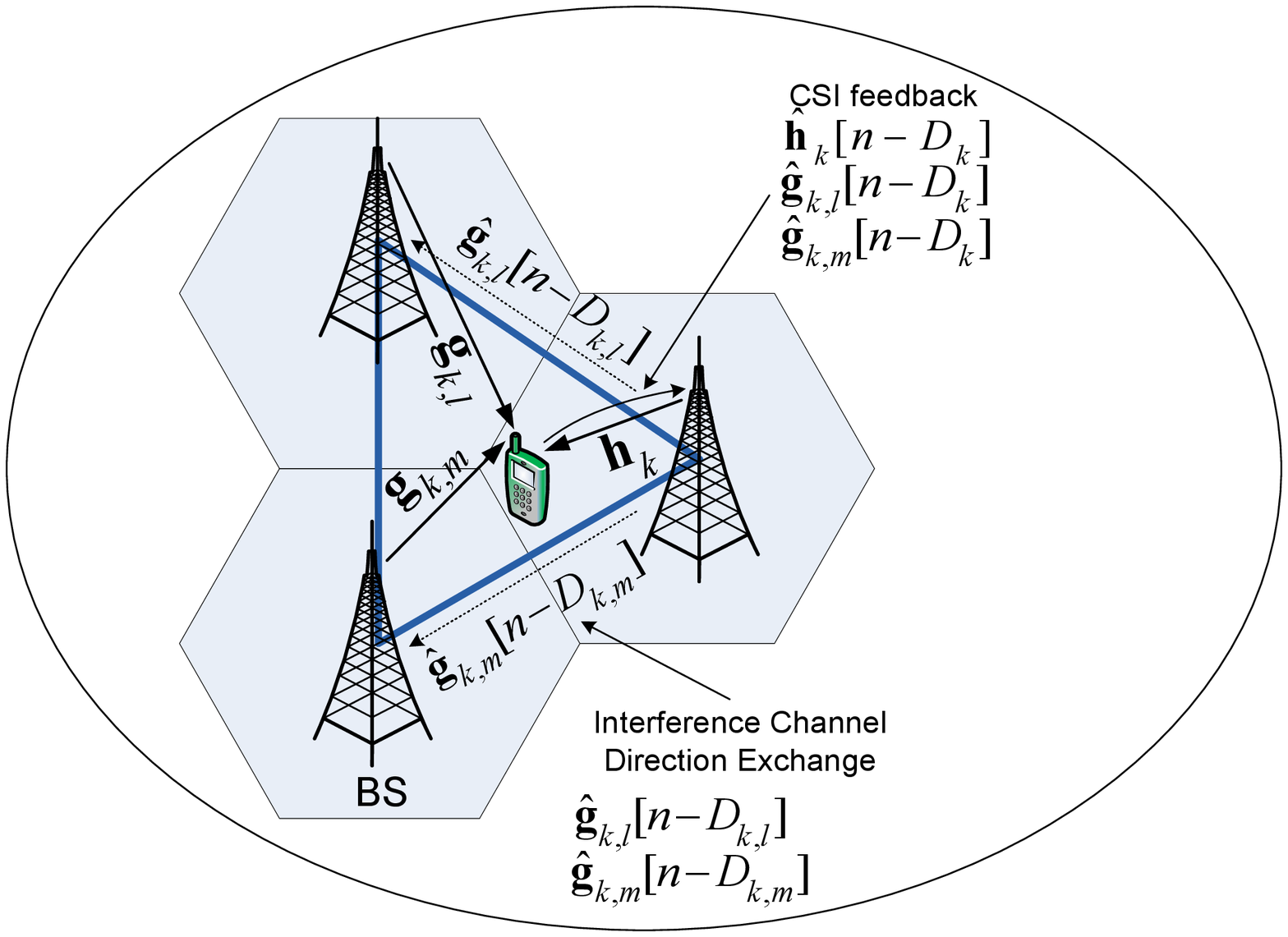}
        \label{fig:sm_carc}
    }
    \subfigure[Split Architecture (SARC)]
    {
        \includegraphics[width = 0.96\columnwidth,height = 2.75in]{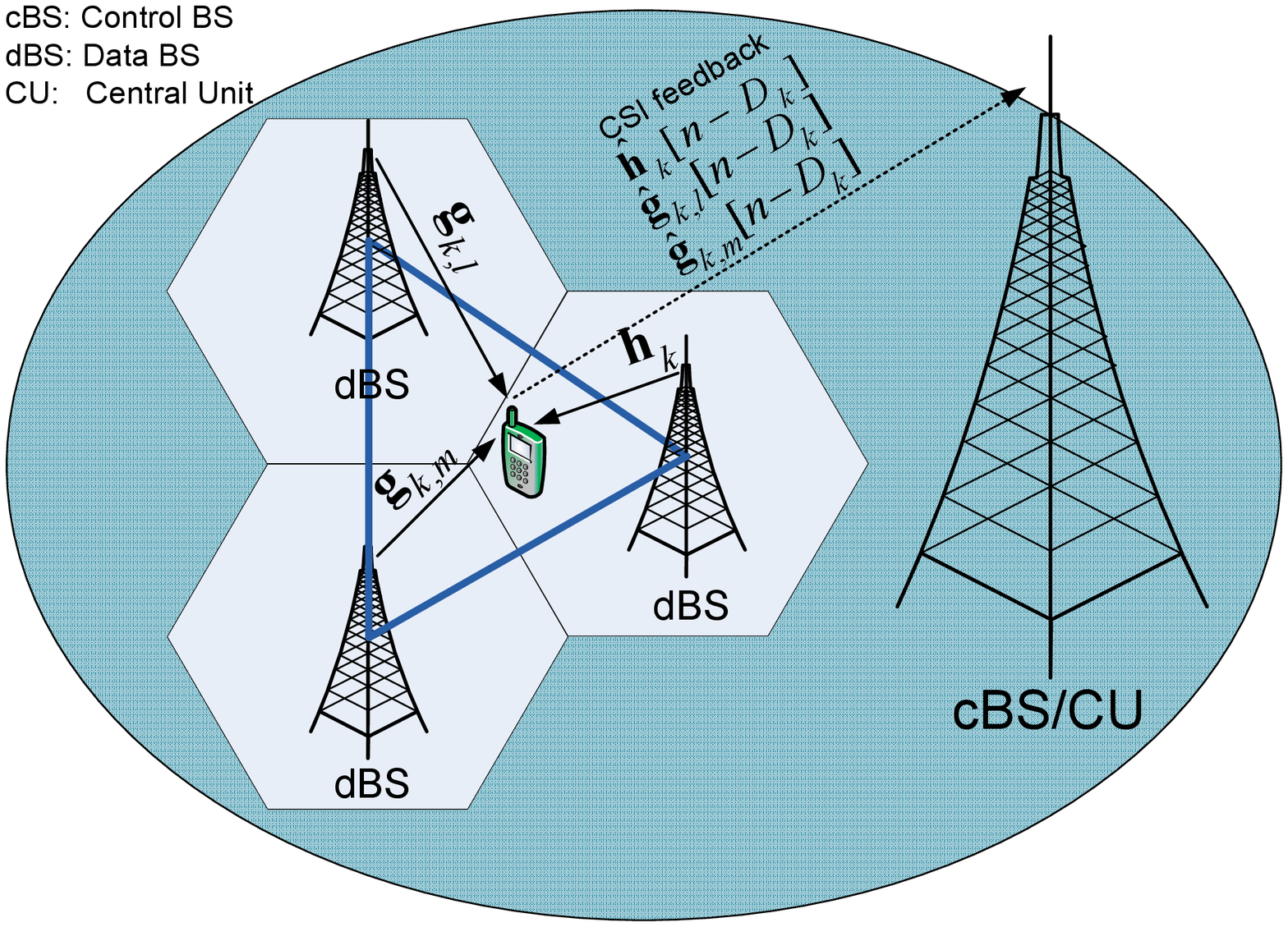}
        \label{fig:sm_sarc}
    }
    \caption{System model for exchange of desired and interfering channels}
    \label{Figure:sm}
\vspace{-3mm}
\end{figure*}

The spectrum used for D2D communication can be either inband or outband. The inband spectrum is considered as licensed cellular band whereas outband spectrum means unlicensed e.g., ISM band. D2D links can operate either in FDD, time division duplex (TDD)\nomenclature{TDD}{Time Division Duplex} \citep{6103908, 7008349, 7073589}, or full duplex mode \citep{6785425,6882650,6952081,7041028}. In FDD mode, two separate frequencies for transmit and receive are required at both nodes of D2D link. This results in under-utilization of spectrum by underlay D2D network. In order to overcome under-utilization of frequencies, TDD mode can be used where single frequency is required for transmit and receive. This comes with more complex transceiver design. The full duplex model allows single frequency without slot sharing (as in TDD mode), however, this can be possible if self-interference due to simultaneous transmission/reception can be canceled. The interested reader is referred to \citep{7041028} for further details. The TDD and full duplex modes have potential advantages of cost-effective transceiver design in small form factor. D2D communication is possible without network, however, it has limited applications as compared to network-assisted direct communication. The capacity and coverage can be enhanced either by overlay or underlay system model. In case of overlay communication, the dedicated spectrum is allocated for D2D network. This can be done by partitioning the available spectrum for cellular and D2D users. In this system model interference management can be relaxed due to allocating dedicated spectrum. However, this model results in low frequency reuse and waste of cellular resources \citep{Overlay15, mumtaz_smart_2014}. A more complex underlay model can be realized where maximum capacity and coverage can be achieved by sharing the same spectrum between cellular and D2D users (full frequency reuse) by incorporating more sophisticated interference management techniques. The interference management comprises either power or RRM depending on uplink or downlink spectrum reuse.

The reader is referred to \citep{5706317} and \citep{6888496} for further details on CoMP and more recent study on CoMP for 5G networks, respectively. For D2D communication, \citep{6805125} and \citep{6970763} provide comprehensive survey and tutorial on the subject.
\section{Cooperation in SARC}\label{sec:coop}
The cooperation in next generation ultra-dense HetNet is indispensable especially when huge D2D links are exploited. In this section, we provide preliminary discussions for possible extension of cooperation framework in SARC. In this context, we first present coordinated beamforming followed by D2D clustering and D2D CoMP in SARC. We further discuss realization of SARC in cloud-RAN architecture, fronthaul/backhaul limitations and possible solution in the form of pro-active caching.

\subsection{Coordinated Beamforming}\label{sec:cb}
In coordinated beamforming, the desired and interfering CSI (ICI) is required at each participating BS of the cooperation set. In conventional multicell CARC, UE measures channel state of serving and neighboring BSs and reports the quantized channel information to the serving BS. The serving BS sorts out ICI and exchanges corresponding interference information to the participating BSs. The participating BSs receive delayed interference information via backhaul and choose appropriate beamformers. In this mechanism, there are two drawbacks. First, the exchange of CSIs between cooperating BSs incurs backhaul delay in addition to the feedback delay from UEs (refer \citep{6666229,4150700} for further details). Secondly, in case the CSI is perturbed (due to quantization effects, noise etc) during exchange via backhaul, the interference at the neighboring cells cannot be perfectly removed resulting in sub-optimal performance \citep{5962731}. In order to highlight these problems, a simple system model of three cells is considered where exchange of desired and interfering channels for SARC and CARC are, respectively, compared in Fig. \ref{Figure:sm}. 
In this figure, the downlink (uplink) desired and interfering channels at UE (BS/dBS) are, respectively $\textbf{h}_{k}$ and $\textbf{g}_{k,x}$ for $x \in \{l,m\}$. The UE normalizes and qunatizes these channels to $\hat{\textbf{h}}_{k}[n]$ and $\hat{\textbf{g}}_{k,x}[n]$, respectively. These channel are fed back by the UE to the serving BS/dBS. The purpose of limited (qunatized) feedback is to send the channel direction to the serving BSs \citep{4150700} where multi-antenna beamforming (single-cell) or CB (multicell) vectors are chosen in such a way that they lie in the null space of interference channel directions \citep{5755206} to achieve inter-cell interference nulling.

The feedback delay associated with CSI is $D_{k}$. Upon receiving the CSI, each BS segregates and forwards ICI to the respective cooperating BSs via backhaul which causes an additional delay $D_{k,x}$ resulting into a total delay of $D_{bh} = D_{k} + D_{k,x}$ where $D_{k,x} \geq D_{k}$. The relation between the current and delayed CSI and ICI is given by Gauss-Markov auto-regressive model \citep{891214} that assumes slowly time varying channels as follows \citep{5755206}:
\begin{align}
\hat{\textbf{h}}_{k}[n] =& \, \eta_{k} \hat{\textbf{h}}_{k} [n-D_{k}]+\sqrt{1-\eta^{2}_{k}}\textbf{e}_{h_{k}}[n],	\IEEEnonumber
\\\hat{\textbf{g}}_{k,x}[n]=& \, \eta_{k,x}\hat{\textbf{g}}_{k,x}[n-D_{k,x}]+\sqrt{1-\eta^{2}_{k,x}}\textbf{e}_{g_{k,x}}[n],
\label{GM_hkgkm_limited}
\end{align}
where $\textbf{e}_{h_{k}}[n]$ and $\textbf{e}_{g_{k,x}}[n]$  are, respectively, desired and interferer channel error vectors distributed as $\mathcal{CN}(0,1)$. The auto-correlation function of desired and interfering channel are $\eta_{k}$ and $\eta_{k,x}$, respectively, defined by the Clarke's auto-correlation model \citep{891214}, \citep{6779222} as:
\begin{align}
\eta_{k} =& b_{0}\, J_{0}\big(2\pi D_{k}f_{d}T_{s}\big),	\IEEEnonumber
\\\eta_{k,x} =& b_{0}\, J_{0}\big(2\pi D_{k,x}f_{d}T_{s}\big),
\label{Correlation}
\end{align}
where $b_{0}$ is the variance of the underlying Gaussian process, $J_{0}(.)$ is the zeroth-order Bessel function of the first kind, $f_{d}$ is the maximum Doppler frequency, and $T_{s}$ is the symbol duration.

Based on above formulation, in Fig. \ref{fig:sm_carc}, it can be seen that each BS exchanges the quantized interference channel between cooperation set. This ICI experiences asymmetric backhaul delay $D_{k,x}$. In such a distributed architecture, the coherent beamforming  can not be achieved and, hence, the benefits of CB can not be fully exploited. However, in SARC, the CSI/ICI is fed back directly to the cBS, therefore $D_{k,x}$ associated with the dBS $x$ is reduced to $D_{x}$. For this case, the auto-correlation function of the interfering channel and corresponding ICI becomes
\begin{align}
\eta_{k,x} =& b_{0}\, J_{0}\big(2\pi D_{k}f_{d}T_{s}\big) = \eta_{k},	\IEEEnonumber
\\\hat{\textbf{g}}_{k,x}[n]=& \, \eta_{k}\hat{\textbf{g}}_{k,x}[n-D_{k}]+\sqrt{1-\eta^{2}_{k}}\textbf{e}_{g_{k,x}}[n],
\label{GM_hkgkm_limited1}
\end{align}

By reducing $\eta_{k,x} = \eta_{k}$ in (\ref{Correlation}) and $D_{k,x} = D_k$ in (\ref{GM_hkgkm_limited}), we can see that, in Fig. \ref{fig:sm_sarc}, the backhaul delay has been eliminated due to direct feedback from the UEs to cBS and hence all beamformers for the participating dBSs can be designed coherently. Although the coherent beamforming can be carried out in SARC, the real problem is to share the beamformers to the corresponding dBSs via asymmetric backhaul links. This problem can be tackled by incorporating centralized timing advance mechanism in cBS CU to allow the participating dBSs adjust the transmission to achieve coherent CB.

The distributed beamforming suits to CARC architecture where CSI/ICI is available in a distributed manner and beamformers are designed at every participating BS. In this case, the perturbation of ICI and corresponding backhaul delay directly effects the performance of CB. However, in SARC, due to inherent centralized ubiquitous coverage, the CSI/ICI from UEs can directly be fed back to cBS. The cBS can act as a CU to design coherent beamformers based on large number of measurement reports. The advantage of this approach is that the backhaul signaling for exchange of interference information and corresponding asymmetric delay can be removed. This approach can further adapt the channel conditions more rapidly since the beamforming does not depend on backhaul delays. In order to address perturbation issue due to exchange of ICI via backhaul links, the availability of global CSI at cBS can be leveraged to jointly design beamforming matrix. Although the exchange of jointly designed beamforming matrix may also be perturbed while exchanging beamformers to the cooperating dBSs via backhaul, the perturbation will effect the overall matrix and results in fair system performance unlike distributed beamformer design in case of CARC.
\subsection{D2D Clustering}\label{sec:Open_Area}
In order to enhance spectral efficiency of cellular systems, intra-cell interference has been tackled in LTE and LTE-A by using OFDMA technology and RRM. Therefore, intra-cell interference is not a problem in such systems, however, inter-cell interference exists for which cooperative communication (CB, and CoMP) has been suggested \citep{marsch_coordinated_2011} to coordinate interference between clusters of BSs and improve cell-edge performance.

In future ultra-dense HetNets, underlay D2D network is being considered as an integral part for rapidly evolving proximal inter-networking. This smallest communication tier reuses the resources of primary users within a cell and hence again generates intra-cell interference which was previously mitigated by OFDMA technology. If we extend the granularity of cooperation at device level and utilize centralized context and CSI (due to separation framework) at cBS CU, we can flexibly control intra/inter-cell interference and hence meet huge capacity gains and spectral efficiency demands of future cellular systems without compromising energy efficiency, and overhead signaling cost (e.g., at air interface or backhaul links). We can further improve these metrics by exploiting self-organized D2D clusters and network controlled D2D communication.

\begin{figure}[!thb]
\centering
\includegraphics[scale = 0.40]{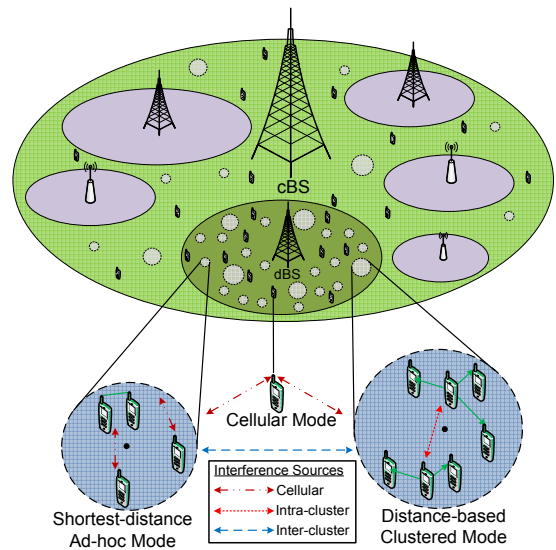}
\caption{D2D clusters and interference sources in SARC.}\label{Figure:distancebasedd2d}
\vspace{-4mm}
\end{figure}

In the following, we consider hierarchical HetNet (i.e., D2D tiers in cBS as well as dBS tiers \citep{7063540}) in SARC and realize D2D communication using channel condition and/or social relationship between nodes as shown in Fig. \ref{Figure:distancebasedd2d}.
\subsubsection{Channel Conditions based D2D clusters}
D2D communication can be realized either in ad-hoc mode or in the form of clusters. In ad-hoc mode, we consider point-to-point links between devices. Since two nodes are allowed to communicate based on shortest distance (reduced path-loss) criterion therefore, in this case, small cooperation radius is required. This mode is feasible for exchange of already cached common information between two devices. However, this mode undermines the potential capacity gain due to the rejection of other nodes that might come in the cooperation radius and request the same common information.
In clustered mode, we consider point-to-multipoint links between devices, therefore, requiring comparatively higher cooperation radius. This mode is feasible for content dissemination. Based on the channel conditions or simple reduced path-loss criterion, one node can be selected by the network to disseminate contents to the requesting nodes. This mode offers higher capacity gain as compared to ad-hoc mode of D2D communication. For cluster regions in ad-hoc and clustered mode, we foresee interference due to:
\begin{itemize}
\item Primary cellular user.
\item Intra-cluster D2D nodes.
\item Inter-cluster D2D nodes.
\end{itemize}

In order to minimize mutual interference between cellular and D2D users, the power optimization at conventional serving BS should consider uplink power control of not only cellular users but also transmit power of near-by D2D nodes. This can be possible if serving BS request near-by D2D nodes to share CSI between the nodes. The CSI may also be used for network-assisted centralized or distributed beamforming to mitigate intra-cluster interference. Similarly, if we incorporate inter-cluster level cooperation, further capacity gains may be envisaged.
\subsubsection{Social network based D2D clusters}\label{ss:snbd2dc}
\begin{figure}[!htb]
\centering
\includegraphics[width = 1\columnwidth]{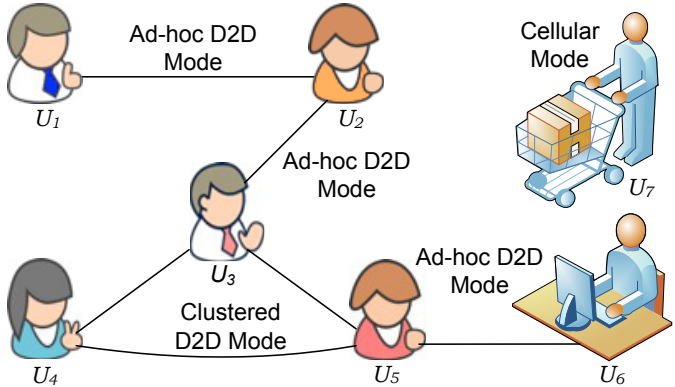}
\caption{Social network offers cellular, ad-hoc, and clustered D2D modes.}\label{Figure:socialnetworkd2d}
\end{figure}

The channel conditions based clustering of D2D nodes is realistic, however it provides overestimated spectral gains due to the assumption that every node has common information to exchange with every other node. In order to assume realistic assumption about common information exchange or content dissemination, social-aware D2D communication should be considered. The social influence of different mobile users may be quantized into different levels of social impact by exploiting the history and logs of each user. For example, some mobile users have limited social influence in terms of assisting the network for content dissemination or offloading and they fit into the category of cellular mode or ad-hoc mode D2D communication. On contrary, many mobile users fall into the category of clustered type D2D communication where they can actively assist the network for exchange of common information and content dissemination. Such social influence may be exploited to model realistic and optimum D2D links/clusters. The exemplary social network for different levels of social interaction is shown in Fig. \ref{Figure:socialnetworkd2d}.

In this figure, D2D link establishment can be done by considering different decision criterion. In this context, different users may be assigned different levels of social influence. The social influence can be calculated using measures of centrality. For example, we can use simple measure of closeness centrality to assign weights to different users in Fig. \ref{Figure:socialnetworkd2d}. The closeness centrality is defined as the shortest distance between a reference node and all other nodes reachable from it \citep{yan2009applying,ni2011degree}. The simple mathematical relation of closeness centrality can be given as:
\begin{equation}
C_c({U_i})=\frac{\big[\sum_{j=1}^{N} d(U_i,U_j)\big]^{-1}}{N-1},
\label{sn}
\end{equation}
where $N$ is the total number of nodes and $d(.)$ is the shortest distance between reference node and all other nodes. Using (\ref{sn}), the closeness centrality\footnote{The closeness centrality has been normalized by the maximum weight in Table \ref{Table:tab5}.} weights can be measured as shown in Table \ref{Table:tab5}.
\begin{table}[h]
\renewcommand{\arraystretch}{1.5}
\centering
\caption{Social influence using closeness centrality.}\label{Table:tab5}
\begin{tcolorbox}[tab4, title=Social Influence,boxrule=0.3mm,top=0.3mm,bottom=0.3mm,left=0.3mm,right=0.3mm,
rightrule=0.3mm]

\begin{tcolorbox}[tab5,tabularx={l|X|X|X|X|X|X|X},title=Closeness Centrality, leftright skip=0.2cm,boxrule=0.25mm,top=0.25mm,bottom=0.25mm,left=0.25mm,right=0.25mm,
rightrule=0.25mm]
\bf Node		& $U_1$	& $U_2$	& $U_3$	& $U_4$	& $U_5$	& $U_6$	& $U_7$	\\ \hline
\bf Closeness	& 0.52	& 0.78	& 1		& 0.78	& 0.78	& 0.56	& 0		
\end{tcolorbox}
\begin{tcolorbox}[tab5, tabularx = {X|X|X}, title=Mode Selection, grow to left by=0.5mm, grow to right by=0.5mm,boxrule=0.25mm,top=0.25mm,bottom=0.25mm,left=0.25mm,right=0.25mm,
rightrule=0.25mm]
\bf{User}			&	\bf Social influence	&	\bf{Mode} 		\\ \hline
$U_7$				&	No			&	Cellular		\\ \hline
$U_1$ and $U_6$		& 	Low			&	Ad-hoc		\\ \hline
$U_2$ :: $U_5$ 		&	High			&	Clustered
\end{tcolorbox}
\end{tcolorbox}
\vspace{-5mm}
\end{table}

According to closeness centrality calculated in Table \ref{Table:tab5}, $U_7$ has no social influence, therefore it is suitable for cellular mode. The users $U_1$ and $U_6$ have low level of social influence and hence they are feasible for ad-hoc mode D2D communication. The users $U_2$, $U_4$ and $U_5$ have slightly higher influence as compared to users $U_1$ and $U_6$ that allows them to be considered for clustered mode D2D communication. In case of ties (e.g., $U_2$, $U_4$, $U_5$), reduced path-loss or better channel conditions based criterion may be used to establish link. User $U_3$ has highest influence which make it suitable for content dissemination in clustered mode of D2D communication.
\subsubsection{Prediction based adaptive D2D clustering}
As mentioned in Sec. \ref{sec:comp_clustering}, the clusters can be static or dynamic where the latter offers more gains as compared to former. The dynamic clustering and cooperation framework is suitable for nomadic users \citep{marsch_coordinated_2011}. Since D2D communication is being evolved for proximity services and inter-networking, dynamic clustering and cooperation framework is very feasible for such type of communication. The dynamic clustering can be extended into self-organized adaptive clustering if the user mobility is predicted. For example, by predicting dwell times of potential D2D users at serving dBS, the required signaling for D2D clustering may be performed in a self-organized manner. Another advantage of this approach is that the prediction of dwell times may allow to tackle ping pong effects and reduce handover cost for switching between cellular and D2D modes. The adaptive clusters can further be optimized by considering mobility patterns along with reduced path-loss, common contents and channel condition criterion.
\subsection{D2D CoMP}\label{sec:d2d_comp}
In previous sub-section, we have presented two modes of D2D communication i.e., ad-hoc and clustered (Fig. \ref{Figure:distancebasedd2d}). In both cases, cooperation framework for multicell BS i.e., CB and CoMP can be realized in SARC for D2D communication. This type of cooperation coupled with common information exchange (ad-hoc mode) or content dissemination (clustered mode) is introduced as D2D CoMP. Since cBS has global context of every node in the coverage area, it can discover nodes for either ad-hoc or clustered mode communication e.g., by localizing nodes and applying shortest distance/reduced path-loss criterion. 

\begin{figure}[!htb]
\centering
\includegraphics[width = 0.96\columnwidth, height = 1.75in]{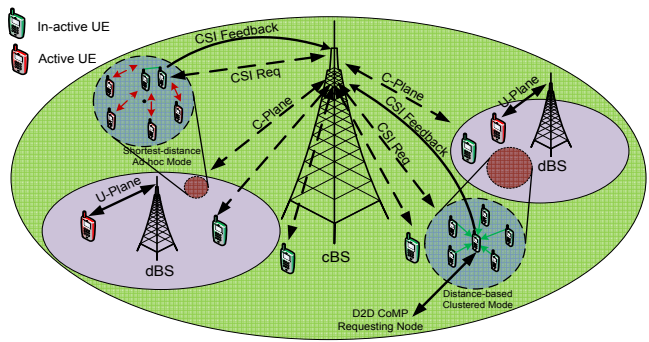}
\caption{D2D CoMP to manage interference in underlay network.}\label{Figure:d2dCoMP}
\vspace{-4mm}
\end{figure}

In order to get CSI between cooperating and requesting nodes, cBS can send a reference signal and request a CSI feedback. Based on RSRP values, one of the node in cooperation cluster may send CSI directly to the cBS. The cBS can use this CSI to design beamformers and share with nodes in cooperation set for proactively cached common information exchange or content dissemination. D2D CoMP in SARC is shown in Fig. \ref{Figure:d2dCoMP}.

In this figure, D2D cooperation regions are shown for ad-hoc and clustered mode D2D CoMP operation. In case of ad-hoc mode, cBS needs to localize and discover an influential partner node\footnote{An influential node can be identified by utilizing the history/context of different nodes and assigning some weight based on the activity of the node e.g., time duration of active sessions, file upload/download frequency etc.} with shortest distance (reduced path-loss) criterion. Once an influential node (containing common information) is identified within proximity of requesting node, cBS can command influential node to send reference signal and subsequently request CSI feedback from the requesting D2D node. For example, in a simple scenario, zero-forcing (ZF) or minimum mean-square-error (MMSE) \citep{6849319} can be used to design precoder to realize CB for ad-hoc mode D2D communication.
 
In case of clustered mode D2D communication, cBS needs to localize a set of influential nodes (known as cooperating nodes in traditional CoMP) that can make cooperation cluster for content dissemination. At this stage, cBS needs to know CSI between requesting and influential nodes. Similar to the ad-hoc mode, cBS can command influential nodes to send reference signal and subsequently request CSI feedback from the requesting node. However, CSI acquisition is more complex as compared to ad-hoc mode due to higher number of distributed influential nodes. Here, we present one strategy to acquire CSI at cBS. In this strategy, cBS will schedule different time slots in a time division multiple access (TDMA) fashion and allocate these slots to the influential nodes. Meanwhile, cBS will command requesting node to acquire time division multiplexed (TDM) reference signals, measure CSI and feedback to the cBS. Once CSI is acquired by the cBS, ZF or MMSE, as mentioned for ad-hoc mode, can be used to design precoders at cBS and shared with influential nodes. The D2D CoMP has potential gains to mitigate interference, however, it comes with the additional cost of higher signaling for CSI acquisition.
\subsection{SARC in Cloud-RAN}\label{sec:cran}
\begin{figure*}[!htb]
\centering
  \includegraphics[width = 0.85\textwidth, height = 1.75in]{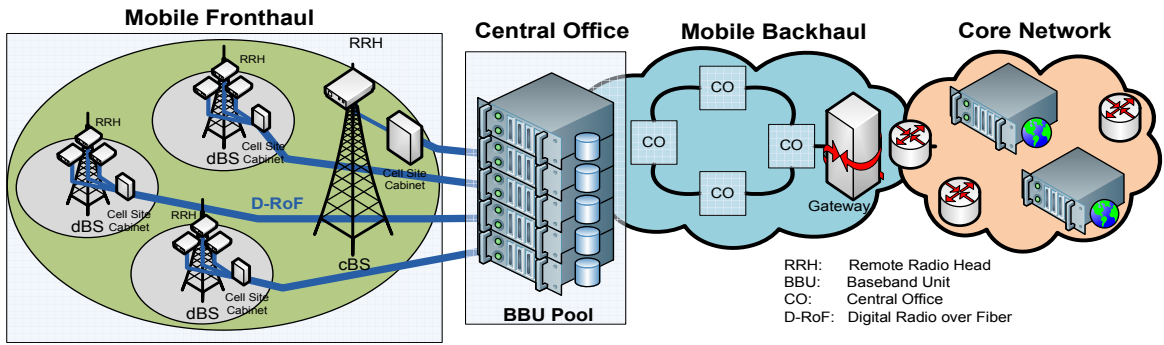}
  \caption{Split C-RAN Architecture}\label{Figure:scran}
\vspace{-5mm}
\end{figure*}

The realization of control and data planes separation has been discussed briefly in \citep{6825019, R1-130566, 6704656} through Carrier Aggregation (CA) and multiple remote radio head (RRH)\nomenclature{RRH}{Remote Radio Head}. Similarly, in \citep{7047300}, the integration of software-defined RAN (SD-RAN) and BCG2 architecture (i.e., decoupled control and data planes) has been suggested to achieve greater benefits and faster realization of both technologies. Motivated by such studies, we present arguments to support SARC in existing cloud RAN (C-RAN)\nomenclature{C-RAN}{Cloud Radio Access Network} architecture. The C-RAN solution comes into two types \citep{CRAN}. The first one is fully centralized where RRH provides radio function and the baseband functions (layer 1, layer 2, etc) are provided by the base band unit (BBU)\nomenclature{BBU}{Base Band Unit}. The second is partially centralized where layer 1 functionality of baseband function is integrated into the RRH. Both C-RAN solutions comprise RRH, the radio function and antennas (located at remote sites as close to the UEs as possible), mobile fronthaul, the fiber link between RRH and BBUs (which can be distributed or centralized at the central office (CO)). In order to realize SARC in C-RAN (SC-RAN)\nomenclature{SC-RAN}{SARC in C-RAN}, some RRHs can be deployed at cBS for ubiquitous coverage and the remaining RRHs for data services. The proposed SC-RAN is shown in Fig. \ref{Figure:scran}.

In this figure, SC-RAN is equivalent to traditional C-RAN with decoupled control and data planes. The BBU stack in CO brings flexibility in C-RAN for joint management of resources and the co-existence of control and data BBUs in SC-RAN can extend this flexibility to share signaling, channel conditions (e.g., CSI), and user data. This results into higher potential to perform joint signal processing e.g., CB and CoMP \citep{CRAN}. The adaptive clustering is more manageable in centralized BBUs in SC-RAN due to global control of the coverage area (cBS BBU). The notion of cell-sleeping can be realized and load balancing, mobility management, and interference management can be accomplished more flexibly with reduced OPEX and higher energy efficiency resulting into future green cellular networks.

The flexibility of realizing SC-RAN comes with the expensive requirement of fronthaul/backhaul links. Since, huge information needs to be exchanged between cooperating dBSs in case of CoMP, high capacity fronthaul/backhaul links are required. In order to address the problem of high capacity backhaul requirements, the distributed caching of contents in femtocells has been proposed in \citep{6195469, 6495773}. These approaches use high storage capacity at femto BS to cache most popular contents and harnessing D2D communication for content delivery. Recently, the backhaul problem in CoMP has been addressed using cache-enabled relays and BSs \citep{6601672, 6665021, 6948325}. All these approaches are based on cache-enabled opportunistic cooperative MIMO (CoMP) framework where a portion of contents are cached at cooperating set of relays or BSs to relax backhaul capacity requirements. Such approaches may be used in SC-RAN, where partially centralized C-RAN (with layer 1 functionality integrated into RRH) can be incorporated so that cache-enabled dBSs can provide high-rate data services in CoMP fashion without requiring huge capacity requirements.
\section{Conclusion}\label{conclusion}
In this article, we outline several performance measures to highlight potential gains and give motivation for evolution of traditional coupled architecture towards control and data planes separation. The different perspectives of energy efficiency, system capacity, interference management and mobility handling are discussed. Since, control and data planes separation approach is in its early stage, little literature exists that addresses some of the performance measures (e.g., \citep{6477646, 6554746} evaluates energy and spectral efficiency). Wherever possible, we provided survey of the approaches proposed for separation architecture; otherwise, we provided our view point for potential advantages and associated complexities in SARC. By considering different scenarios from the perspective of outlined performance measures, it is revealed that there is a huge potential for capacity and energy efficiency enhancements by separating control and data planes. Moreover, the SARC provides flexibility in mobility management at the cost of more complex signaling network. The second part of the article provides background for cooperation framework for interference management in multicell environment. It is emphasized that there are several potential advantages of sending CSI to the cBS and exploiting pro-active caching to realize backhaul relaxed CB and CoMP for interference management in future ultra-dense cellular environment.

Another perspective of cooperation has been presented where cooperation means assisting network for common information exchange or content dissemination between near-by devices in the form of ad-hoc or clustered mode direct communication. D2D CoMP has been introduced where conventional cooperation framework has been suggested to handle intra-cell interference. Due to ubiquitous coverage in SARC, centralized cBS offers more flexibility in CSI acquisition and corresponding beamforming for CB and CoMP operation. The centralized cBS also offers higher degree of freedom to predict nodes for content sharing and it can even be combined with network pro-active caching and adaptive clustering for self-organized D2D communication. 

Motivated by the control and data planes separation framework, in the following, we outline the lessons learned and several potential research directions in this area:
\begin{itemize}
  \item Energy efficiency is the most important aspect of future cellular systems. Among many approaches mentioned in Sec. \ref{ee} (e.g., BS switch-off, smart grid, renewable energy sources), dynamic BS switch-off mechanism can play an important role in realizing green cellular communication. The inherent drawback of coverage holes (due to BS switch-off techniques) and more interference (due to increased transmit power in cell range expansion) does not exist in SARC due to ubiquitous coverage. Some of the research studies (e.g., \citep{6477646, 6554746,7037473}) investigated the potential gains in energy efficiency due to control and data planes separation. In order to investigate full energy efficiency gains, the realistic power consumption models are required. For such models, existing approaches for traditional architecture can be investigated followed by more advanced and sophisticated energy management techniques for SARC.
  \item The higher spectrum and more bandwidth are envisioned to ensure capacity requirements of future cellular networks. In this context, mm-Wave spectrum and carrier aggregation are potential candidates for next generation cellular networks. A lot of research is being conducted to investigate feasibility of mm-Wave spectrum. Designing new channel models for dual connectivity (i.e., mm-Wave for data plane and lower frequency for control plane) has a lot of research potential that can lead towards communication in SARC.
  \item For current HetNet, intra-cell interference does not exist and inter-cell interference management has been standardized. In future ultra-dense networks, intra-cell interference will again be a problem due to underlay systems e.g., D2D communication. In order to overcome this interference, existing techniques of CB and CoMP can be extended at device level (D2D) and the backhaul limitations can be be complemented by exploiting pro-active caching techniques (e.g., \citep{6601672, 6665021, 6948325}). 
  \item In future ultra-dense environment, cells at mm-Wave spectrum will have spot beam coverage. This results in huge capacity enhancements which can further be leveraged by harnessing D2D cooperation for content sharing or content dissemination.
  \item Mobility management is flexible due to higher degree of freedom in SARC. However, this comes at the price of complex signaling network in SARC. The control plane design will be more complex due to more tiers (underlay networks). In this context, lot of research endeavors are required to realize seamless handovers and higher coverage probability while ensuring QoS requirements of each user.
\end{itemize}
\balance
\section*{Acknowledgements}
This work was made possible by NPRP grant No. 5-1047-2437 from Qatar National Research Fund (a member of the Qatar Foundation). The statements made herein are solely the responsibility of the authors. We would also like to acknowledge the support of the University of Surrey 5GIC (http://www.surrey.ac.uk/5gic) members for this work.
\balance
\Urlmuskip=0mu plus 1mu\relax
\bibliographystyle{IEEEtran}
\bibliography{Cooperative_Networks_____1}

\begin{thebibliography}{100}
\providecommand{\url}[1]{#1}
\csname url@samestyle\endcsname
\providecommand{\newblock}{\relax}
\providecommand{\bibinfo}[2]{#2}
\providecommand{\BIBentrySTDinterwordspacing}{\spaceskip=0pt\relax}
\providecommand{\BIBentryALTinterwordstretchfactor}{4}
\providecommand{\BIBentryALTinterwordspacing}{\spaceskip=\fontdimen2\font plus
\BIBentryALTinterwordstretchfactor\fontdimen3\font minus
  \fontdimen4\font\relax}
\providecommand{\BIBforeignlanguage}[2]{{%
\expandafter\ifx\csname l@#1\endcsname\relax
\typeout{** WARNING: IEEEtran.bst: No hyphenation pattern has been}%
\typeout{** loaded for the language `#1'. Using the pattern for}%
\typeout{** the default language instead.}%
\else
\language=\csname l@#1\endcsname
\fi
#2}}
\providecommand{\BIBdecl}{\relax}
\BIBdecl

\bibitem{Cisco}
S.~Cass, ``{IP Traffic in 2017: 1.4 Zettabytes},'' Feb. 2014, {Accessed:
  2015-06-02}.

\bibitem{UMTS_2011}
\BIBentryALTinterwordspacing
``{UMTS Forum Report: Mobile Traffic Forecasts: 2010-2020},'' Report 44, May
  2011, {Accessed: 2015-06-02}. [Online]. Available:
  \url{http://www.umts-forum.org/component/option,com_docman/task,doc_download/gid,2348/Itemid,213/}
\BIBentrySTDinterwordspacing

\bibitem{6692781}
A.~Osseiran \emph{et~al.}, ``{The Foundation of the Mobile and Wireless
  Communications System for 2020 and Beyond: Challenges, Enablers and
  Technology Solutions},'' in \emph{Proc. IEEE Veh. Tech. Conf. (VTC)}, June
  2013, pp. 1--5.

\bibitem{Ericsson_2011}
Ericsson, ``{More than 50 billion connected devices},'' White Paper, February
  2011, {Accessed: 2015-06-02}.

\bibitem{6171992}
J.~Andrews, H.~Claussen, M.~Dohler, S.~Rangan, and M.~Reed, ``Femtocells: Past,
  present, and future,'' \emph{IEEE J. Sel. Areas Commun.,}, vol.~30, no.~3,
  pp. 497--508, April 2012.

\bibitem{5677351}
A.~Bianzino, C.~Chaudet, D.~Rossi, and J.~Rougier, ``{A Survey of Green
  Networking Research},'' \emph{IEEE Commun. Surveys \& Tutorials}, vol.~14,
  no.~1, pp. 3--20, First 2012.

\bibitem{wu_green_2012}
J.~Wu, S.~Rangan, and H.~Zhang, Eds., \emph{\BIBforeignlanguage{English}{{Green
  Communications: Theoretical Fundamentals, Algorithms and Applications}}},
  1st~ed.\hskip 1em plus 0.5em minus 0.4em\relax Boca Raton, FL: CRC Press,
  Sep. 2012.

\bibitem{scott_matthews_planning_2010}
H.~Scott~Matthews \emph{et~al.}, ``\BIBforeignlanguage{en}{{Planning
  energy-efficient and eco-sustainable telecommunications networks}},''
  \emph{\BIBforeignlanguage{en}{Bell Labs Technical Journal}}, vol.~15, no.~1,
  pp. 215--236, Jun. 2010, {Accessed: 2015-06-02}.

\bibitem{5978416}
A.~Fehske, G.~Fettweis, J.~Malmodin, and G.~Biczok, ``The global footprint of
  mobile communications: The ecological and economic perspective,'' \emph{IEEE
  Commun. Mag.}, vol.~49, no.~8, pp. 55--62, August 2011.

\bibitem{6848019}
C.~Gao \emph{et~al.}, ``{Relax, but Do Not Sleep: A new perspective on Green
  Wireless Networking},'' in \emph{Proc. IEEE Conference on Computer
  Communications (INFOCOM)}, April 2014, pp. 907--915.

\bibitem{6525595}
M.~Shakir \emph{et~al.}, ``{Green heterogeneous small-cell networks: toward
  reducing the CO$_2$ emissions of mobile communications industry using uplink
  power adaptation},'' \emph{IEEE Commun. Mag.}, vol.~51, no.~6, pp. 52--61,
  June 2013.

\bibitem{4205092}
A.~Dejonghe \emph{et~al.}, ``{Green Reconfigurable Radio Systems},'' \emph{IEEE
  Signal Process. Mag.}, vol.~24, no.~3, pp. 90--101, May 2007.

\bibitem{6100924}
G.~Rittenhouse, S.~Goyal, D.~Neilson, and S.~Samuel, ``Sustainable
  telecommunications,'' in \emph{Technical Symposium at ITU Telecom World (ITU
  WT)}, Oct 2011, pp. 19--23.

\bibitem{ABI}
{ABIresearch}, ``{Market Gets Primed to Rollout Half a Million Outdoor Small
  Cells in 2013},'' January 2013, {Accessed: 2015-06-02}.

\bibitem{4117538}
P.~Marsch, S.~Khattak, and G.~Fettweis, ``A framework for determining realistic
  capacity bounds for distributed antenna systems,'' in \emph{IEEE Information
  Theory Workshop (ITW)}, Oct 2006, pp. 571--575.

\bibitem{3gpp.36.819}
3GPP, ``{Coordinated Multi-point Operation for LTE Physical Layer Aspects},''
  {3rd Generation Partnership Project (3GPP)}, TS {36.819 V11.1.0}, Dec. 2011.

\bibitem{marsch_coordinated_2011}
P.~Marsch and G.~P. Fettweis, Eds., \emph{{Coordinated Multi-point in Mobile
  Communications: From Theory to Practice}}.\hskip 1em plus 0.5em minus
  0.4em\relax Cambridge ; New York: Cambridge University Press, 2011.

\bibitem{4657145}
S.~Parkvall \emph{et~al.}, ``{LTE-Advanced - Evolving LTE towards
  IMT-Advanced},'' in \emph{Proc. IEEE Veh. Tech. Conf. (VTC)}, Sept 2008, pp.
  1--5.

\bibitem{greentouch}
\BIBentryALTinterwordspacing
``{GreenTouch-Project},'' {Accessed: 2015-06-02}. [Online]. Available:
  \url{http://www.greentouch.org/index.php?page=mobile_networks_working_group_projects}
\BIBentrySTDinterwordspacing

\bibitem{METIS}
\BIBentryALTinterwordspacing
``{METIS} 2020,'' {Accessed: 2015-06-02}. [Online]. Available:
  \url{https://www.metis2020.com}
\BIBentrySTDinterwordspacing

\bibitem{MiWEBA}
\BIBentryALTinterwordspacing
``{MiWEBA - Project},'' {Accessed: 2015-06-02}. [Online]. Available:
  \url{http://www.miweba.eu/?page_id=80}
\BIBentrySTDinterwordspacing

\bibitem{6468982}
E.~Ternon, Z.~Bharucha, and H.~Taoka, ``{A Feasibility Study for the Detached
  Cell Concept},'' in \emph{Proc. Intern. ITG Conf. on Sys. Comm. \& Coding
  (SCC)}, Jan 2013, pp. 1--5.

\bibitem{6152217}
A.~Capone, A.~Fonseca~dos Santos, I.~Filippini, and B.~Gloss, ``{Looking beyond
  green cellular networks},'' in \emph{Annual Conf. on Wireless On-demand Netw.
  Sys. \& Services (WONS)}, 2012, pp. 127--130.

\bibitem{6515050}
X.~Xu, G.~He, S.~Zhang, Y.~Chen, and S.~Xu, ``{On Functionality Separation for
  Green Mobile Networks: Concept Study over LTE},'' \emph{IEEE Commun. Mag.},
  vol.~51, no.~5, pp. 82--90, 2013.

\bibitem{6056691}
G.~Auer \emph{et~al.}, ``{How much energy is needed to run a wireless
  network?}'' \emph{IEEE Wireless Commun.}, vol.~18, no.~5, pp. 40--49, October
  2011.

\bibitem{4448824}
J.~Louhi, ``{Energy efficiency of modern cellular base stations},'' in
  \emph{Intern. Telecommunications Energy Conf. (INTELEC)}, 2007, pp. 475--476.

\bibitem{6673363}
M.~Olsson \emph{et~al.}, ``{5GrEEn: Towards Green 5G mobile networks},'' in
  \emph{IEEE Intern. Conf. on Wireless \& Mobile Comput., Netw. and Comm.
  (WiMob)}, Oct 2013, pp. 212--216.

\bibitem{zhao2013software}
T.~Zhao \emph{et~al.}, ``{Software defined radio implementation of signaling
  splitting in hyper-cellular network},'' in \emph{Proc. of the second workshop
  on Software radio implementation forum}.\hskip 1em plus 0.5em minus
  0.4em\relax ACM, 2013, pp. 81--84.

\bibitem{6702534}
H.~Ali-Ahmad \emph{et~al.}, ``{An SDN-Based Network Architecture for Extremely
  Dense Wireless Networks},'' in \emph{IEEE SDN for Future Netw. \& Services
  (SDN4FNS)}, Nov 2013, pp. 1--7.

\bibitem{6825019}
A.~Zakrzewska, D.~Lopez-Perez, S.~Kucera, and H.~Claussen, ``{Dual connectivity
  in LTE HetNets with split control- and user-plane},'' in \emph{Proc. IEEE
  Global Telecommun. Conf. (GLOBECOM)}, Dec 2013, pp. 391--396.

\bibitem{6477646}
H.~Ishii, Y.~Kishiyama, and H.~Takahashi, ``{A novel architecture for LTE-B
  :C-plane/U-plane split and Phantom Cell concept},'' in \emph{Proc. IEEE
  Global Telecommun. Conf. (GLOBECOM)}, Dec 2012, pp. 624--630.

\bibitem{6554746}
S.~Mukherjee and H.~Ishii, ``{Energy Efficiency in the Phantom Cell enhanced
  Local Area architecture},'' in \emph{IEEE Wireless Communications and
  Networking Conference (WCNC)}, April 2013, pp. 1267--1272.

\bibitem{7067574}
Y.~Okumura, ``{5G mobile radio access system using SHF/EHF bands},'' in
  \emph{Asia-Pacific Microw. Conf. (APMC)}, Nov 2014, pp. 908--910.

\bibitem{6848637}
S.~Suyama, J.~Shen, A.~Benjebbour, Y.~Kishiyama, and Y.~Okumura, ``{Super high
  bit rate radio access technologies for small cells using higher frequency
  bands},'' in \emph{IEEE MTT-S Intern. Microw. Symposium (IMS)}, June 2014,
  pp. 1--4.

\bibitem{wwrf2009}
\BIBentryALTinterwordspacing
K.~E.~S. WWRF, L.~Sorensen, ``User scenarios 2020,'' July 2009, {Accessed:
  2015-06-02}. [Online]. Available:
  \url{http://www.wireless-world-research.org}
\BIBentrySTDinterwordspacing

\bibitem{tafazolli2006technologies}
R.~Tafazolli, \emph{Technologies for the Wireless Future: Wireless World
  Research Forum (WWRF)}.\hskip 1em plus 0.5em minus 0.4em\relax John Wiley \&
  Sons, 2006.

\bibitem{Huawei}
Y.~Q. Bian and D.~Rao, ``{Small Cells Big Opportunities},'' Huawei, February
  2014, {Accessed: 2015-06-02}.

\bibitem{FierceMobileIT}
\BIBentryALTinterwordspacing
{ Fred Donovan}, ``{Infonetics: In-building, outdoor small cells to handle
  quarter of mobile traffic by 2016},'' FierceMobileIT, January 2013,
  {Accessed: 2015-06-02}. [Online]. Available:
  \url{http://www.fiercemobileit.com}
\BIBentrySTDinterwordspacing

\bibitem{SCFORUM}
{ Small Cell Forum}, ``{Small cells – what’s the big idea?}'' February
  2012, {Accessed: 2015-06-02}.

\bibitem{5722322}
D.~Ferling \emph{et~al.}, ``{Energy efficiency approaches for radio nodes},''
  in \emph{Future Netw. \& Mobile Summit (FutureNetworkSummit)}, June 2010, pp.
  1--9.

\bibitem{5621969}
L.~Correia \emph{et~al.}, ``{Challenges and enabling technologies for energy
  aware mobile radio networks},'' \emph{IEEE Commun. Mag.}, vol.~48, no.~11,
  pp. 66--72, 2010.

\bibitem{6600717}
H.~Holtkamp, G.~Auer, V.~Giannini, and H.~Haas, ``{A Parameterized Base Station
  Power Model},'' \emph{IEEE Commun. Lett. (COMML)}, vol.~17, no.~11, pp.
  2033--2035, November 2013.

\bibitem{6629715}
M.~Herlich and H.~Karl, ``{Energy-Efficient Assignment of User Equipment to
  Cooperative Base Stations},'' in \emph{Intern. Symposium in Wireless Comm.
  Sys. (ISWCS)}, Aug 2013, pp. 1--5.

\bibitem{7037473}
S.~Zhang, J.~Wu, J.~Gong, S.~Zhou, and Z.~Niu, ``{Energy-optimal probabilistic
  base station sleeping under a separation network architecture},'' in
  \emph{Proc. IEEE Global Telecommun. Conf. (GLOBECOM)}, Dec 2014, pp.
  4239--4244.

\bibitem{5208045}
M.~Marsan, L.~Chiaraviglio, D.~Ciullo, and M.~Meo, ``{Optimal Energy Savings in
  Cellular Access Networks},'' in \emph{Proc. IEEE International Conference on
  Communications Workshop (ICC)}, June 2009, pp. 1--5.

\bibitem{5300273}
L.~Chiaraviglio, D.~Ciullo, M.~Meo, and M.~Marsan, ``{Energy-efficient
  management of UMTS access networks},'' in \emph{Intern. Teletraffic Congress
  (ITC)}, Sept 2009, pp. 1--8.

\bibitem{5683654}
E.~Oh and B.~Krishnamachari, ``{Energy Savings through Dynamic Base Station
  Switching in Cellular Wireless Access Networks},'' in \emph{Proc. IEEE Global
  Telecommun. Conf. (GLOBECOM)}, Dec 2010, pp. 1--5.

\bibitem{5360741}
A.~Fehske, F.~Richter, and G.~Fettweis, ``{Energy Efficiency Improvements
  through Micro Sites in Cellular Mobile Radio Networks},'' in \emph{Proc. IEEE
  Global Telecommun. Conf. (GLOBECOM),}, Nov 2009, pp. 1--5.

\bibitem{rost201011}
P.~Rost and G.~Fettweis, ``{Green communications in cellular networks with
  fixed relay nodes},'' \emph{Cooperative Cellular Wireless Networks}, p. 300,
  2010.

\bibitem{6489498}
E.~Oh, K.~Son, and B.~Krishnamachari, ``{Dynamic Base Station Switching-On/Off
  Strategies for Green Cellular Networks},'' \emph{IEEE Trans. Wireless
  Commun.}, vol.~12, no.~5, pp. 2126--2136, May 2013.

\bibitem{5992823}
K.~Son, H.~Kim, Y.~Yi, and B.~Krishnamachari, ``{Base Station Operation and
  User Association Mechanisms for Energy-Delay Tradeoffs in Green Cellular
  Networks},'' \emph{IEEE J. Sel. Areas Commun.,}, vol.~29, no.~8, pp.
  1525--1536, 2011.

\bibitem{6731020}
H.~Al~Haj~Hassan, L.~Nuaymi, and A.~Pelov, ``{Renewable energy in cellular
  networks: A survey},'' in \emph{IEEE Online Conf. on Green Commun.
  (GreenCom)}, Oct 2013, pp. 1--7.

\bibitem{6290252}
L.-C. Wang and S.~Rangapillai, ``{A survey on green 5G cellular networks},'' in
  \emph{Intern. Conf. on Signal Process. \& Commun. (SPCOM)}, July 2012, pp.
  1--5.

\bibitem{NTT2004}
S.~K.~E. NTT~DOCOMO, Sorensen~L, ``{FOMA base station using solar and wind
  power},'' July 2004, {Accessed: 2015-06-02}.

\bibitem{WWF2010}
\BIBentryALTinterwordspacing
Z.~P. Yang~Tianjian, Hu~Yiwen and D.~Pamlin, ``{Low carbon telecommunications
  solutions in china},'' July 2010, {Accessed: 2015-06-02}. [Online].
  Available: \url{http://www.pamlin.net/other_documents/China\ Mobile\
  report-summary-en.pdf}
\BIBentrySTDinterwordspacing

\bibitem{6102353}
S.~Bu, F.~Yu, and P.~Liu, ``{A game-theoretical decision-making scheme for
  electricity retailers in the smart grid with demand-side management},'' in
  \emph{IEEE Intern. Conf. on Smart Grid Comm. (SmartGridComm)}, Oct 2011, pp.
  387--391.

\bibitem{6364971}
S.~Bu, F.~Yu, Y.~Cai, and P.~Liu, ``{Energy efficient cellular networks with
  CoMP communications and smart grid},'' in \emph{Proc. IEEE International
  Conference on Communications (ICC)}, June 2012, pp. 5921--5925.

\bibitem{6210335}
S.~Bu, F.~Yu, Y.~Cai, and X.~Liu, ``{When the Smart Grid Meets Energy-Efficient
  Communications: Green Wireless Cellular Networks Powered by the Smart
  Grid},'' \emph{IEEE Trans. Wireless Commun.}, vol.~11, no.~8, pp. 3014--3024,
  August 2012.

\bibitem{5706319}
S.~Liu, J.~Wu, C.~H. Koh, and V.~Lau, ``{A 25 Gb/s(/km$^2$) urban wireless
  network beyond IMT-advanced},'' \emph{IEEE Commun. Mag.}, vol.~49, no.~2, pp.
  122--129, February 2011.

\bibitem{IMT}
\BIBentryALTinterwordspacing
``Estimated spectrum bandwidth requirements for the future development of
  {IMT}-2000 and {IMT}-{Advanced},'' {Accessed: 2015-06-02}. [Online].
  Available: \url{http://www.itu.int/pub/R-REP-M.2078}
\BIBentrySTDinterwordspacing

\bibitem{6881734}
R.~Zhang, Z.~Zheng, M.~Wang, X.~Shen, and L.-L. Xie, ``{Equivalent Capacity in
  Carrier Aggregation-Based LTE-A Systems: A Probabilistic Analysis},''
  \emph{IEEE Trans. Wireless Commun.}, vol.~13, no.~11, pp. 6444--6460, Nov
  2014.

\bibitem{DOCOMO}
``{5G Radio Access: Requirements, Concept and Technologies},'' pp. 1--13,
  {Accessed: 2015-06-02}.

\bibitem{Artemis}
\BIBentryALTinterwordspacing
``{ARTEMIS},'' {Accessed: 2015-06-02}. [Online]. Available:
  \url{http://www.artemis.com/pcell}
\BIBentrySTDinterwordspacing

\bibitem{6732923}
S.~Rangan, T.~Rappaport, and E.~Erkip, ``{Millimeter-Wave Cellular Wireless
  Networks: Potentials and Challenges},'' \emph{Proceedings of the IEEE}, vol.
  102, no.~3, pp. 366--385, March 2014.

\bibitem{4457895}
R.~Daniels and R.~Heath, ``{60 GHz wireless communications: emerging
  requirements and design recommendations},'' \emph{IEEE Veh. Technol. Mag.},
  vol.~2, no.~3, pp. 41--50, September 2007.

\bibitem{1491267}
M.~Marcus and B.~Pattan, ``{Millimeter wave propagation; spectrum management
  implications},'' \emph{IEEE Microw. Mag.}, vol.~6, no.~2, pp. 54--62, June
  2005.

\bibitem{6824752}
J.~Andrews \emph{et~al.}, ``{What Will 5G Be?}'' \emph{IEEE J. Sel. Areas
  Commun.,}, vol.~32, no.~6, pp. 1065--1082, June 2014.

\bibitem{6134693}
M.~Jacob \emph{et~al.}, ``{Diffraction in mm and Sub-mm Wave Indoor Propagation
  Channels},'' \emph{IEEE Trans. Microw. Theory Techn.}, vol.~60, no.~3, pp.
  833--844, March 2012.

\bibitem{6253227}
M.~Kyro, V.~Kolmonen, and P.~Vainikainen, ``{Experimental Propagation Channel
  Characterization of mm-Wave Radio Links in Urban Scenarios},'' \emph{IEEE
  Antennas Wireless Propag. Lett.}, vol.~11, pp. 865--868, 2012.

\bibitem{6824972}
S.~Larew, T.~Thomas, M.~Cudak, and A.~Ghosh, ``{Air interface design and ray
  tracing study for 5G millimeter wave communications},'' in \emph{Proc. IEEE
  Global Telecommun. Conf. (GLOBECOM)}, Dec 2013, pp. 117--122.

\bibitem{6515173}
T.~Rappaport \emph{et~al.}, ``{Millimeter Wave Mobile Communications for 5G
  Cellular: It Will Work!}'' \emph{IEEE Access}, vol.~1, pp. 335--349, 2013.

\bibitem{6736750}
W.~Roh \emph{et~al.}, ``{Millimeter-wave beamforming as an enabling technology
  for 5G cellular communications: theoretical feasibility and prototype
  results},'' \emph{IEEE Commun. Mag.}, vol.~52, no.~2, pp. 106--113, February
  2014.

\bibitem{4784727}
T.~Janevski, ``{5G Mobile Phone Concept},'' in \emph{IEEE Consumer
  Communications and Networking Conference (CCNC)}, Jan 2009, pp. 1--2.

\bibitem{tudzarov_protocols_2011}
A.~Tudzarov and T.~Janevski, ``\BIBforeignlanguage{en}{{Protocols and
  Algorithms for the Next Generation 5G Mobile Systems}},''
  \emph{\BIBforeignlanguage{en}{Network Protocols and Algorithms}}, vol.~3,
  no.~1, pp. 94--114, Jun. 2011.

\bibitem{Nanocore}
``{5G the Nanocore},'' pp. 1--24, {Accessed: 2015-06-02}.

\bibitem{5441362}
M.~Rahman and H.~Yanikomeroglu, ``{Enhancing cell-edge performance: a downlink
  dynamic interference avoidance scheme with inter-cell coordination},''
  \emph{IEEE Trans. Wireless Commun.}, vol.~9, no.~4, pp. 1414--1425, April
  2010.

\bibitem{6392819}
C.~Kosta, B.~Hunt, A.~Quddus, and R.~Tafazolli, ``{On Interference Avoidance
  Through Inter-Cell Interference Coordination (ICIC) Based on OFDMA Mobile
  Systems},'' \emph{IEEE Commun. Surveys \& Tutorials}, vol.~15, no.~3, pp.
  973--995, Third 2013.

\bibitem{4907410}
G.~Boudreau \emph{et~al.}, ``{Interference coordination and cancellation for 4G
  networks},'' \emph{IEEE Commun. Mag.}, vol.~47, no.~4, pp. 74--81, April
  2009.

\bibitem{5506110}
Z.~Xie and B.~Walke, ``{Frequency Reuse Techniques for Attaining Both Coverage
  and High Spectral Efficiency in OFDMA Cellular Systems},'' in \emph{IEEE
  Wireless Communications and Networking Conference (WCNC)}, April 2010, pp.
  1--6.

\bibitem{5450256}
L.~Dong, Z.~Song, L.~Wenxin, and W.~Wenbo, ``{A frequency reuse partitioning
  scheme with successive interference cancellation for OFDMA uplink
  transmission},'' in \emph{IEEE International Symposium on Personal, Indoor
  and Mobile Radio Communications (PIMRC)}, September 2009, pp. 1362--1366.

\bibitem{6475212}
S.~Deb, P.~Monogioudis, J.~Miernik, and J.~Seymour, ``{Algorithms for Enhanced
  Inter-Cell Interference Coordination (eICIC) in LTE HetNets},''
  \emph{IEEE/ACM Trans. Netw.}, vol.~22, no.~1, pp. 137--150, February 2014.

\bibitem{D_Lopez2011}
D.~Lopez-Perez \emph{et~al.}, ``{Enhanced inter-cell interference coordination
  challenges in heterogeneous networks},'' \emph{IEEE Trans. Wireless Commun.},
  vol.~18, no.~3, pp. 22--30, Jun 2011.

\bibitem{6725662}
A.~Liu, V.~Lau, L.~Ruan, J.~Chen, and D.~Xiao, ``{Hierarchical Radio Resource
  Optimization for Heterogeneous Networks With Enhanced Inter-Cell Interference
  Coordination (eICIC)},'' \emph{IEEE Trans. Signal Process.}, vol.~62, no.~7,
  pp. 1684--1693, April 2014.

\bibitem{7022987}
M.~Karabacak, D.~Wang, H.~Ishii, and H.~Arslan, ``{Mobility Performance of
  Macrocell-Assisted Small Cells in Manhattan Model},'' in \emph{Proc. IEEE
  Veh. Tech. Conf. (VTC)}, May 2014, pp. 1--5.

\bibitem{7059729}
C.-H. Lee and Z.-S. Syu, ``{Handover Analysis of Macro-Assisted Small Cell
  Networks},'' in \emph{IEEE International Conference on Internet of
  Things(iThings), and IEEE Green Computing and Communications (GreenCom) and
  IEEE Cyber, Physical and Social Computing (CPSCom)}, September 2014, pp.
  604--609.

\bibitem{7063430}
S.~Kuklinski, Y.~Li, and K.~T. Dinh, ``{Handover management in SDN-based mobile
  networks},'' in \emph{Proc. IEEE Global Telecommun. Conf. (GLOBECOM)}, Dec
  2014, pp. 194--200.

\bibitem{7063379}
L.~Valtulina, M.~Karimzadeh, G.~Karagiannis, G.~Heijenk, and A.~Pras,
  ``{Performance evaluation of a SDN/OpenFlow-based Distributed Mobility
  Management (DMM) approach in virtualized LTE systems},'' in \emph{Proc. IEEE
  Global Telecommun. Conf. (GLOBECOM)}, Dec 2014, pp. 18--23.

\bibitem{6994333}
D.~Kreutz \emph{et~al.}, ``{Software-Defined Networking: A Comprehensive
  Survey},'' \emph{Proceedings of the IEEE}, vol. 103, no.~1, pp. 14--76, Jan
  2015.

\bibitem{6807723}
A.~Sniady and J.~Soler, ``{LTE for Railways: Impact on Performance of ETCS
  Railway Signaling},'' \emph{IEEE Veh. Technol. Mag.}, vol.~9, no.~2, pp.
  69--77, June 2014.

\bibitem{6710300}
L.~Yan and X.~Fang, ``{Decoupled wireless network architecture for high-speed
  railway},'' in \emph{IEEE International Workshop on High Mobility Wireless
  Communications (HMWC)}, Nov 2013, pp. 96--100.

\bibitem{choi_standards_2014}
H.~Y. Choi, Y.~Song, and Y.-K. Kim, ``{Standards of Future Railway Wireless
  Communication in Korea}.''\hskip 1em plus 0.5em minus 0.4em\relax Recent
  Advances in Computer Engineering, Communications and Information Technology,
  2014, {Accessed: 2015-06-02}.

\bibitem{3gpp.36.331}
3GPP, ``{Evolved Universal Terrestrial Radio Access, Radio Resource Control},''
  {3rd Generation Partnership Project (3GPP)}, TS {36.331 V11.5.0 (2013-09)}.

\bibitem{ahn2011wireless}
C.~W. Ahn and J.-H. Lee, ``{Wireless cooperative communication: a survey},'' in
  \emph{International Conference on Ubiquitous Information Management and
  Communication (ICUIMC)}, 2011, p.~78.

\bibitem{rubinstein2006survey}
M.~G. Rubinstein, I.~M. Moraes, M.~E.~M. Campista, L.~H. M.~K. Costa, and O.~C.
  M.~B. Duarte, ``\BIBforeignlanguage{en}{{A Survey on Wireless Ad Hoc
  Networks}},'' in \emph{\BIBforeignlanguage{en}{{Mobile and Wireless
  Communication Networks}}}, G.~Pujolle, Ed.\hskip 1em plus 0.5em minus
  0.4em\relax Springer US, 2006, no. 211, pp. 1--33.

\bibitem{1194444}
B.~Hochwald and S.~ten Brink, ``{Achieving near-capacity on a multiple-antenna
  channel},'' \emph{IEEE Trans. Commun.}, vol.~51, no.~3, pp. 389--399, March
  2003.

\bibitem{738086}
P.~Wolniansky, G.~Foschini, G.~Golden, and R.~Valenzuela, ``{V-BLAST: an
  architecture for realizing very high data rates over the rich-scattering
  wireless channel},'' in \emph{URSI International Symposium on Signals,
  Systems, and Electronics}, Sep 1998, pp. 295--300.

\bibitem{774855}
X.~Wang and H.~Poor, ``{Iterative (turbo) soft interference cancellation and
  decoding for coded CDMA},'' \emph{IEEE Trans. Commun.}, vol.~47, no.~7, pp.
  1046--1061, Jul 1999.

\bibitem{1271237}
H.~Dai, A.~Molisch, and H.~Poor, ``{Downlink capacity of interference-limited
  MIMO systems with joint detection},'' \emph{IEEE Trans. Wireless Commun.},
  vol.~3, no.~2, pp. 442--453, March 2004.

\bibitem{1207369}
G.~Caire and S.~Shamai, ``{On the achievable throughput of a multiantenna
  Gaussian broadcast channel},'' \emph{IEEE Trans. Inf. Theory}, vol.~49,
  no.~7, pp. 1691--1706, July 2003.

\bibitem{1291726}
N.~Jindal, S.~Vishwanath, and A.~Goldsmith, ``{On the duality of Gaussian
  multiple-access and broadcast channels},'' \emph{IEEE Trans. Inf. Theory},
  vol.~50, no.~5, pp. 768--783, May 2004.

\bibitem{4203115}
W.~Yu and T.~Lan, ``{Transmitter Optimization for the Multi-Antenna Downlink
  With Per-Antenna Power Constraints},'' \emph{IEEE Trans. Signal Process.},
  vol.~55, no.~6, pp. 2646--2660, June 2007.

\bibitem{garavaglia_adaptive_2014}
A.~Garavaglia, R.~Weber, M.~Schulist, and S.~Brueck, ``{Adaptive Cell
  Clustering in a Multi-cluster Environment},'' U.S. Patent {US} 8\,639\,256
  B2, Jan., 2014.

\bibitem{5675775}
M.~Corson \emph{et~al.}, ``{Toward proximity-aware internetworking},''
  \emph{IEEE Wireless Commun.}, vol.~17, no.~6, pp. 26--33, December 2010.

\bibitem{balraj_lte_2012}
S.~Balraj, ``{LTE Direct Overview},'' {Accessed: 2015-06-02}.

\bibitem{MOTO}
\BIBentryALTinterwordspacing
``{Radio Access and Spectrum {FP}7 - Future Networks Cluster},'' {Accessed:
  2015-06-02}. [Online]. Available:
  \url{http://www.ict-ras.eu/index.php/ras-projects/moto}
\BIBentrySTDinterwordspacing

\bibitem{6881261}
E.~Bastug, M.~Bennis, and M.~Debbah, ``Social and spatial proactive caching for
  mobile data offloading,'' in \emph{Proc. IEEE International Conference on
  Communications Workshop (ICC)}, June 2014, pp. 581--586.

\bibitem{6103908}
Y.~Li, T.~Zhou, J.~Xu, Z.~Li, and H.~Wang, ``{Adaptive TDD UL/DL slot
  utilization for cellular controlled D2D communications},'' in \emph{Global
  Mobile Congress (GMC)}, Oct 2011, pp. 1--6.

\bibitem{7008349}
H.~Sun, M.~Sheng, M.~Wildemeersch, and T.~Quek, ``{Modeling of D2D enhanced
  two-tier dynamic TDD heterogeneous cellular networks},'' in \emph{IEEE/CIC
  International Conference on Communications in China (ICCC)}, Oct 2014, pp.
  609--614.

\bibitem{7073589}
H.~Sun, M.~Wildemeersch, M.~Sheng, and T.~Quek, ``{D2D Enhanced Heterogeneous
  Cellular Networks with Dynamic TDD},'' \emph{IEEE Trans. Wireless Commun.},
  vol.~PP, no.~99, pp. 1--1, 2015.

\bibitem{6785425}
S.~Kim and W.~Stark, ``{Full Duplex Device to Device Communication in Cellular
  Networks},'' in \emph{International Conference on Computing, Networking and
  Communications (ICNC)}, Feb 2014, pp. 721--725.

\bibitem{6882650}
S.~Ali, N.~Rajatheva, and M.~Latva-Aho, ``{Full Duplex Device-to-Device
  Communication in Cellular Networks},'' in \emph{European Conference on
  Networks and Communications (EuCNC)}, June 2014, pp. 1--5.

\bibitem{6952081}
K.~Hemachandra, N.~Rajatheva, and M.~Latva-Aho, ``{Sum-rate Analysis for
  Full-duplex Underlay Device-to-device Networks},'' in \emph{IEEE Wireless
  Communications and Networking Conference (WCNC)}, April 2014, pp. 514--519.

\bibitem{7041028}
S.~Ali, A.~Ghazanfari, N.~Rajatheva, and M.~Latva-aho, ``{Effect of residual of
  self-interference in performance of full-duplex D2D communication},'' in
  \emph{International Conference on 5G for Ubiquitous Connectivity (5GU)}, Nov
  2014, pp. 46--51.

\bibitem{Overlay15}
C.~B.~Das, ``{A Study on Device To Device Communication in Wireless Mobile
  Network},'' \emph{International Journal of Modern Communication Technologies
  \& Research}, vol.~3, no.~3, Mar. 2015.

\bibitem{mumtaz_smart_2014}
\BIBentryALTinterwordspacing
S.~Mumtaz and J.~Rodriguez, Eds., \emph{\BIBforeignlanguage{en}{{Smart Device
  to Smart Device Communication}}}.\hskip 1em plus 0.5em minus 0.4em\relax
  Cham: Springer International Publishing, 2014. [Online]. Available:
  \url{http://link.springer.com/10.1007/978-3-319-04963-2}
\BIBentrySTDinterwordspacing

\bibitem{5706317}
R.~Irmer \emph{et~al.}, ``{Coordinated multipoint: Concepts, performance, and
  field trial results},'' \emph{IEEE Commun. Mag.}, vol.~49, no.~2, pp.
  102--111, February 2011.

\bibitem{6888496}
S.~Schwarz and M.~Rupp, ``{Exploring Coordinated Multipoint Beamforming
  Strategies for 5G Cellular},'' \emph{IEEE Access}, vol.~2, pp. 930--946,
  2014.

\bibitem{6805125}
A.~Asadi, Q.~Wang, and V.~Mancuso, ``{A Survey on Device-to-Device
  Communication in Cellular Networks},'' \emph{IEEE Commun. Surveys \&
  Tutorials}, vol.~16, no.~4, pp. 1801--1819, Fourthquarter 2014.

\bibitem{6970763}
J.~Liu, N.~Kato, J.~Ma, and N.~Kadowaki, ``{Device-to-Device Communication in
  LTE-Advanced Networks: A Survey},'' \emph{IEEE Commun. Surveys \& Tutorials},
  vol.~PP, no.~99, pp. 1--1, 2014.

\bibitem{6666229}
B.~E. Godana and D.~Gesbert, ``{Coordinated beamforming in multicell networks
  with Channel State Information exchange delays},'' in \emph{IEEE
  International Symposium on Personal, Indoor and Mobile Radio Communications
  (PIMRC)}, Sept 2013, pp. 713--718.

\bibitem{4150700}
K.~Huang, B.~Mondal, R.~Heath, and J.~Andrews, ``{CTH07-1: Effect of Feedback
  Delay on Multi-Antenna Limited Feedback for Temporally-Correlated
  Channels},'' in \emph{Proc. IEEE Global Telecommun. Conf. (GLOBECOM)}, Nov
  2006, pp. 1--5.

\bibitem{5962731}
R.~Bhagavatula and R.~Heath, ``Impact of delayed limited feedback on the
  sum-rate of intercell interference nulling,'' in \emph{Proc. IEEE
  International Conference on Communications (ICC)}, June 2011, pp. 1--5.

\bibitem{5755206}
------, ``{Adaptive Bit Partitioning for Multicell Intercell Interference
  Nulling With Delayed Limited Feedback},'' \emph{IEEE Trans. Signal Process.},
  vol.~59, no.~8, pp. 3824--3836, Aug 2011.

\bibitem{891214}
C.~Tan and N.~Beaulieu, ``{On first-order Markov modeling for the Rayleigh
  fading channel},'' \emph{IEEE Trans. Commun.}, vol.~48, no.~12, pp.
  2032--2040, Dec 2000.

\bibitem{6779222}
R.~Clarke, ``{A statistical theory of mobile-radio reception},'' \emph{Bell
  System Technical Journal, The}, vol.~47, no.~6, pp. 957--1000, July 1968.

\bibitem{7063540}
H.~Mustafa \emph{et~al.}, ``{Spectral efficiency improvements in HetNets by
  exploiting device-to-device communications},'' in \emph{Proc. IEEE Global
  Telecommun. Conf. (GLOBECOM)}, Dec 2014, pp. 857--862.

\bibitem{yan2009applying}
E.~Yan and Y.~Ding, ``{Applying centrality measures to impact analysis: A
  coauthorship network analysis},'' \emph{Journal of the American Society for
  Information Science and Technology}, vol.~60, no.~10, pp. 2107--2118, 2009.

\bibitem{ni2011degree}
C.~Ni, C.~Sugimoto, and J.~Jiang, ``{Degree, Closeness, and Betweenness:
  Application of group centrality measurements to explore macro-disciplinary
  evolution diachronically},'' in \emph{Proc. International Conference of the
  International Society for Scientometrics \& Informetrics (ISSI)}, 2011, pp.
  1--13.

\bibitem{6849319}
S.~Mumtaz, K.~Saidul~Huq, and J.~Rodriguez, ``{Coordinated paradigm for D2D
  communications},'' in \emph{IEEE Conference on Computer Communications
  Workshops (INFOCOM WKSHPS)}, April 2014, pp. 718--723.

\bibitem{R1-130566}
Ericsson, ``{Physical Layer Aspects of Dual Connectivity},'' {St. Julian's,
  Malta, 3GPP Standard Contribution R1-130566}, Feb. 2013.

\bibitem{6704656}
K.~Sakaguchi \emph{et~al.}, ``{Cloud cooperated heterogeneous cellular
  networks},'' in \emph{International Symposium on Intelligent Signal
  Processing and Communication Systems (ISPACS)}, Nov 2013, pp. 787--791.

\bibitem{7047300}
Z.~Zaidi, V.~Friderikos, and M.~Imran, ``{Future RAN Architecture: SD-RAN
  Through a General-Purpose Processing Platform},'' \emph{IEEE Veh. Technol.
  Mag.}, vol.~10, no.~1, pp. 52--60, March 2015.

\bibitem{CRAN}
``{C-RAN the Road Towards Green RAN},'' pp. 1--44, {Accessed: 2015-06-02}.

\bibitem{6195469}
N.~Golrezaei, K.~Shanmugam, A.~Dimakis, A.~Molisch, and G.~Caire,
  ``{FemtoCaching: Wireless video content delivery through distributed caching
  helpers},'' in \emph{Proc. IEEE Conference on Computer Communications
  (INFOCOM)}, March 2012, pp. 1107--1115.

\bibitem{6495773}
N.~Golrezaei, A.~Molisch, A.~Dimakis, and G.~Caire, ``{Femtocaching and
  device-to-device collaboration: A new architecture for wireless video
  distribution},'' \emph{IEEE Commun. Mag.}, vol.~51, no.~4, pp. 142--149,
  April 2013.

\bibitem{6601672}
A.~Liu and V.~Lau, ``{Mixed-Timescale Precoding and Cache Control in Cached
  MIMO Interference Network},'' \emph{IEEE Trans. Signal Process.}, vol.~61,
  no.~24, pp. 6320--6332, Dec 2013.

\bibitem{6665021}
------, ``{Cache-Enabled Opportunistic Cooperative MIMO for Video Streaming in
  Wireless Systems},'' \emph{IEEE Trans. Signal Process.}, vol.~62, no.~2, pp.
  390--402, Jan 2014.

\bibitem{6948325}
------, ``{Exploiting Base Station Caching in MIMO Cellular Networks:
  Opportunistic Cooperation for Video Streaming},'' \emph{IEEE Trans. Signal
  Process.}, vol.~63, no.~1, pp. 57--69, Jan 2015.

\end{thebibliography}

\end{document}